__________________________________________________________________________
\documentstyle[12pt]{article}

\begin{document}
\begin{centering}
\large\bf The induced representations of Brauer algebra\\
and the Clebsch-Gordan coefficients of SO(n)
\vskip .6truecm
\normalsize Feng Pan$^{1\dagger}$, Shihai Dong$^{2}$ and J. P. Draayer$^{1}$
\vskip .2cm
\noindent{\small\it $^{1}$Department of Physics {\normalsize\&} Astronomy, 
Louisiana State University,}\\
{\small\it Baton Rouge, LA 70803-4001}\\
\vskip .1cm
{\small\it $^{2}$Department of Physics, Liaoning Normal University, 
Dalian 116029, P. R. China}\\
\vskip 1truecm
{\bf Abstract}\\
\end{centering}
\vskip .5cm
   \normalsize Induced representations of Brauer algebra $D_{f}(n)$ from 
   $S_{f_{1}}\times S_{f_{2}}$ with
$f_{1}+f_{2}=f$ are discussed. The induction coefficients (IDCs) or the 
outer-product reduction
coefficients (ORCs) of  $S_{f_{1}}\times S_{f_{2}}\uparrow D_{f}(n)$ 
with $f\leq 4$ up
to a normalization factor are derived by using the linear equation method. 
Weyl tableaus for
the corresponding Gel'fand basis of SO(n) are defined. The assimilation 
method for obtaining
CG coefficients of SO(n) in the Gel'fand basis for no modification rule 
involved 
couplings  from IDCs of Brauer algebra are proposed. Some 
isoscalar factors of $SO(n)\supset SO(n-1)$ for the resulting irrep
$[\lambda_{1},~\lambda_{2},~
\lambda_{3},~\lambda_{4},\dot{0}]$
with $\sum\limits_{i=1}^{4}\lambda_{i}\leq 4$ are tabulated.
\vskip 4.0cm
\noindent {-----------------------------------------}\\
\noindent $^{\dagger}$On leave from Department of Physics,
Liaoning Normal Univ., Dalian 116029, P. R.~~China

\newpage

\begin{centering}
{\large 1. Introduction}\\
\end{centering}
\vskip .4truecm
   Clebsch-Gordan Coefficients (CGCs) are of importance in many physical 
problems. Besides
those$^{[1,2]}$ of SO(3) and SO(4), which  were discussed extensively and 
expressed in various forms,
Isoscalar Factors (ISFs) of $SO(n)\supset SO(n-1)$, which can be used to 
evaluate CGCs of
SO(n) in its canonical basis  according to the Racah factorization 
lemma,$^{[3]}$ were derived analytically
for the coupling $[l_{1},~\dot{0}] \times [l_{2},~\dot{0}]$ to 
$[L_{1},L_{2},\dot{0}]$ by using
substitution group technique.$^{[4]}$  The ISFS for the coupling 
$[l_{1},l_{2},~\dot{0}]\times
[l_{3},~\dot{0}]$ to $[L_{1},L_{2},\dot{0}]$  of some special cases were 
derived by using
group chain transformation method.$^{[5]}$  A  special class of 
multiplicity-free $O(n)\supset O(n-1)$
isoscalar factors were derived in the Gel'fand basis in [6]. Very recently,
isoscalar factors of $O(n)\supset O(n-1)$ for  the coupling 
$[l_{1},l_{2},l_{3},\dot{0}]\times 
[1,\dot{0}]$ have been derived by using the irreducible tensor basis 
method.$^{[7]}$ However, unlike those of SU(n) case, CGCs of SO(n) in its
canonical basis or ISFs of
$SO(n)\supset SO(n-1)$, generally, are rank $n$ dependent, which makes 
it difficult to derive them for
general case analytically. On the other hand,  analytical expressions of 
the CGCs in most cases
are very complicated and difficult to be used in  applications. 
Therefore,  tables of CGCs
or ISFs, if available, are more convenient in practical use.

\vskip .3cm
   In this paper, we will outline a procedure for  deriving CGCs of 
$SO(n)$ in its
canonical basis from Induction coefficients (IDCs) of $S_{f_{1}}\times 
S_{f_{2}}\uparrow D_{f}(n)$.
 Brauer algebra $D_{f}(n)$, which are similar to the group algebra of
symmetric group $S_{f}$ related to the decomposition  of $f$-rank tensors 
of the general linear
group GL(n), are the centralizer algebra of the  orthogonal group O(n) 
or the symplectic group
Sp(2m) when $n=-2m$. More precisely, if G is the orthogonal group O(n) 
or the symplectic group
Sp(2m), the corresponding centralizer algebra $B_{f}(G)$ are quotients 
of Brauer's $D_{f}(n)$ and 
$D_{f}(-2m)$, respectively. Hence, duality relation between $D_{f}(n)$ 
and O(n) or Sp(2m) is quite
the same as the Schur-Weyl duality relation  between $S_{f}$ and GL(n). 
Irreducible representations
of $D_{f}(n)$ in the standard  basis, i.e.  the basis adapted to the 
chain $D_{f}(n)\supset 
D_{f-1}(n)\supset\cdots\supset D_{2}(n)$, have been constructed by using 
the induced representation
and linear equation method,$^{[8]}$ and  more elaborately by Leduc and 
Ram using the so-called ribbon
Hopf algebra approach.$^{[9]}$ Racah  coefficients of O(n) and Sp(2m) 
were successfully derived from Subduction
coefficients (SDCs) of $D_{f}(n)$ by  using the Brauer-Schur-Weyl duality 
relation.$^{[10]}$ A new simple Young
diagrammatic method for Kronecker  products of O(n) and Sp(2m) was also 
formulated,$^{[11]}$ 
which, actually, is  based on the  induced representation
theory of Brauer algebra discussed in this paper.   
\vskip .3cm
   In Sec. 2, we will briefly review  irreps of $D_{f}(n)$ in the 
standard basis. Then, induced
representations of  $S_{f_{1}}\times  S_{f_{2}}\uparrow D_{f}(n)$ with 
$f_{1}+f_{2}=f$ will be defined.
In Sec. 3, based on the linear equation method, which has been proved
effective in evaluating IDCs, SDCs of Hecke   algebra and SDCs of Brauer 
algebra,
a procedure  for the evaluation of IDCs of $D_{f}(n)$  will be outlined. 
In Sec. 4, Weyl tabeaux for SO(n) in its
canonical basis will be defined. Then, a general  procedure for 
evaluating CGCs of SO(n) in its
canonical basis for no modification rule involved couplings
will be outlined. Finally, some analytical  expressions for the ISFs of 
$SO(n)\supset
SO(n-1)$ for the  resulting irrep 
$[\lambda_{1},~\lambda_{2},~\lambda_{3},~\lambda_{4},\dot{0}]$
with $\sum_{i=1}^{4}\lambda_{i}\leq 4$
will be tabulated in Sec. 5.
\vskip .5cm
\begin{centering}
{\large 2. Brauer algebra and its outer-product basis}\\
\end{centering}
\vskip .3cm
   Brauer algebra $D_{f}(n)$ is defined algebraically by $2f-2$ generators
$\{ g_{1},~g_{2},~\cdots,~g_{f-1},~e_{1},$\\
$e_{2},~\cdots,~e_{f-1}\}$  satisfying the following relations$^{[12]}$

$$g_{i}g_{i+1}g_{i}= 
g_{i+1}g_{i}g_{i+1}, ~~~~g_{i}g_{j}=g_{j}g_{i},~~\vert i-j\vert \geq 
2,\eqno(1a)$$

$$e_{i}g_{i}=e_{i},~~~~e_{i}g_{i-1}e_{i}=e_{i}.\eqno(1b)$$

\noindent Using the above defined relations and by drawing pictures
of link diagrams,$^{[12-13]}$ one can also derive other useful
ones. For example,

$$e_{i}e_{j}=e_{j}e_{i},~~~\vert i-j\vert\geq 2,$$

$$e_{i}^{2}=ne_{i},$$

$$(g_{i}-1)^{2}(g_{i}+1)=0.\eqno(2)$$
\vskip .3cm
  We assume that the base field is ${\bf C}$. 
The star operation,  a conjugate linear map $\dagger$, on $D_{f}(n)$ is 
defined by

$$g^{\dagger}_{i}=g_{i},~~~ e^{\dagger}_{i}=e_{i}~~ {\rm for}~~ 
i=1,2,\cdots,f-1,\eqno(3)$$

\noindent which are necessary in defining orthonormal basis
of $D_{f}(n)$. 
\vskip .3cm 
 It is easy to see  that $\{ g_{1},~g_{2},~\cdots,~g_{f-1}\}$
generate a subalgebra ${\bf C}S_{f}$, which is isomorphic to the group
algebra of the symmetric group; that is, $D_{f}(n)\supset {\bf C}S_{f}$.
The properties of $D_{f}(n)$ have been discussed in [12-13].
Based on these results, it is known that $D_{f}(n)$ is semisimple, i. e. it
is a direct sum of full matrix algebra over $\bf C$, when $n$ is not an integer
or is an integer with $n\geq f-1$, otherwise $D_{f}(n)$ is no longer semisimple.
 In the following, we assume that $n$ is an integer with $n\geq f-1$. In this 
case $D_{f}(n)$ is semisimple. Irreducible
representations of $D_{f}(n)$ can be denoted by a Young diagram with
$f,~f-2,~f-4,~\cdots,~1$ or $0$ boxes. An irrep of $D_{f}(n)$ with
$f-2k$ boxes is denoted as $[\lambda ]_{f-2k}$. The branching rule
of $D_{f}(n)\downarrow D_{f-1}(n)$ is

$$[\lambda ]_{f-2k} =\oplus_{[\mu]\leftrightarrow [\lambda ]}[\mu ],$$ 

\noindent where $[\mu ]$ runs through all the diagrams obtained by removing or
(if $[\lambda ]$ contains less than $f$ boxes) adding a box to $[\lambda ]$.
Hence, the basis vectors of $D_{f}(n)$ in the standard basis can be denoted
by

$$\left\vert
\matrix{
~~~~~[\lambda]_{f-2k} &D_{f}(n)\cr
[\mu ] &D_{f-1}(n)\cr
\vdots &\vdots\cr
[p] &D_{f-p+1}(n)\cr
[\nu ] &D_{f-p}(n)\cr}
\right) =
\left\vert\matrix{
[\lambda ]_{f-2k}\cr
[\mu ]\cr
\vdots\cr
[p]\cr
Y^{[\nu]}_{M}\cr}\right),\eqno(4)$$ 

\noindent where $[\nu ]$  is identical to the same irrep of $S_{f-p}$, 
$Y^{[\nu ]}_{M}$ 
is a standard Young tableau,  and $M$ can be understood either as the 
Yamanouchi
symbols or indices of the basis  vectors in the so-called decreasing page 
order
of the Yamanouchi symbols. Irreps  in the standard basis given by (4) 
were given in
[8] for $f\leq 5$. Higher dimensional  results can also derived by using 
the method given in [8] or by using Leduc and Ram's formulae.$^{[9]}$
\vskip .3cm
   In order to study CGCs of SO(n), we  need consider induced 
representations of
Brauer algebra, $S_{f_{1}}\times  S_{f_{2}}\uparrow D_{f}(n)$ with 
$f_{1}+f_{2}=f$, for the  outer-products

$$[\lambda_{1}]\times [\lambda_{2}]\uparrow\sum_{\lambda}\{\lambda_{1}
\lambda_{2}\lambda\}[\lambda ],\eqno(5)$$

\noindent where $\{\lambda_{1}\lambda_{2}\lambda\}$ is number of occurrence
of irrep $[\lambda ]$ in the outer-product $[\lambda_{1}]\times[\lambda_{2}]$.
The standard basis vectors of $[\lambda_{1}]_{f_{1}}$,
and $[\lambda_{2}]_{f_{2}}$ for $D_{f_{1}}(n)$ and $D_{f_{2}}(n)$,
which are the same as those for $S_{f_{1}}$ and $S_{f_{2}}$, can be
denoted by $\vert Y^{[\lambda_{1}]}_{m_{1}}(\omega^{0}_{1})>$, and
$\vert Y^{[\lambda_{2}]}_{m_{2}}(\omega^{0}_{2})>$, respectively, where

$$(\omega^{0}_{1})=(1,~2,~\cdots,~f_{1}),~~~~(\omega_{2}^{0})
=(f_{1}+1,~f_{1}+2,~\cdots,~f_{1}+f_{2})\eqno(6)$$

\noindent are indices in the standard tableaus $Y^{[\lambda_{1}]}_{m_{1}}$
and $Y^{[\lambda_{2}]}_{m_{2}}$, respectively. The product of the two basis
vectors is denoted by

$$\vert Y^{[\lambda_{1}]}_{m_{1}},~ 
Y^{[\lambda_{2}]}_{m_{2}},~(\omega_{1}^{0}),~(\omega_{2}^{0})>
\equiv \vert Y^{[\lambda_{1}]}_{m_{1}}(\omega_{1}^{0})>\vert 
Y^{[\lambda_{2}]}_{m_{2}}(\omega^{0}_{2})>,\eqno(7)$$

\noindent which is called primitive uncoupled basis vector.
\vskip .3cm
   When $n$ is a positive integer, we can use tensor products of the rank-1
 unit tensor operator of $O(n)$ to construct the basis  of $D_{f}(n)$  in 
the standard basis
 explicitly through so-called cabling [8]. In this case the indices 
$1,2,\cdots, f$ are used to 
 distinguish tensor operators from different spaces. We also need a set 
of the corresponding
 indices $i_{1},i_{2},\cdots, i_{f}$ to label the tensor components which 
can be taken as $n$ different values, namely

$$T^{1}_{i_{1}}T^{2}_{i_{2}}\cdots T^{f}_{i_{f}} \equiv T^{12\cdots f}
_{i_{1}i_{2}\cdots i_{f}}.\eqno(8)$$

\noindent The actions of $g_{i}$ and $e_{i}$ on (8) are given by

$$g_{i}T^{1~2\cdots i~i+1\cdots f}_{j_{1}j_{2}\cdots j_{i}j_{i+1}\cdots j_{f}}
=T^{1~2\cdots i+1~i\cdots f}_{j_{1}j_{2}\cdots j_{i}j_{i+1}\cdots j_{f}},$$

$$e_{i}T^{1~2\cdots i~i+1\cdots f}_{j_{1}j_{2}\cdots j_{i}j_{i+1}\cdots j_{f}}
 =\delta_{j_{i},j_{i+1}}\sum_{j}{}^{(s)}T^{1~2\cdots i~i+1\cdots 
f}_{j_{1}j_{2}\cdots j,j\cdots j_{f}},\eqno(9a)$$

\noindent where the sum, $\sum^{(s)}$, on the right hand means

$$\sum_{j}{}^{(s)}T^{12}_{jj}=\sum_{j\notin SO(2)}T^{12}_{jj}-(T^{1~~2}_{\alpha_{2}~-a_{2}}+
T^{~1~~2}_{-\alpha_{2}~\alpha_{2}}).\eqno(9b)$$

\noindent In order to discuss couplings of $SO(n)$ in the canonical 
 basis, i.e. the basis adapted to $SO(n)\supset SO(n-1)\supset 
SO(n-2)\supset \cdots\supset SO(2)$,
the rank-1 $SO(n)$ tensor components are
 classified according to the $SO(n)\supset SO(n-1)$ reduction. Namely, 
the tensor components of $j$ for rank-1 tensor
$T^{[1]}_{j}$ of $SO(n)$  are labeled by $j=\pm \alpha_{2},\alpha_{3}$,
$\cdots \alpha_{n}$.
The minus sign introduced in (9b) are consistent with the Condon-Shortly
phase convention$^{[1]}$ for CGCs of $SO(3)$. 
 We assume that $\{T^{1~2\cdots f}_{j_{1}j_{2}\cdots j_{f}}\}$
spans a orthonormal inner product space, namely

 $$\left( T^{1^{\prime}~2^{\prime}\cdots 
f^{\prime}}_{j^{\prime}_{1}j^{\prime}_{2}\cdots j^{\prime}_{f}},
 T^{1~2\cdots f}_{j_{1}j_{2}\cdots 
j_{f}}\right)=\prod\delta_{ii^{\prime}}\delta_{j_{i}j_{i}^{\prime}}.\eqno(10)$$

 \noindent Then, the primitive uncoupled basis vectors given by (7) can 
be expressed in terms of 
 these $T$ operators. For example, $S_{1}\times S_{1}$ basis vector can 
be expressed as

$$\vert 1>=\vert 1,2>= T^{1}_{i_{1}}T^{2}_{i_{2}}.\eqno(11)$$

 \noindent Other uncoupled basis vectors can be obtained by acting 
$g_{1}$, and $e_{1}$, respectively, on (11).

 $$\vert 2>=g_{1}\vert 1>=T^{2}_{i_{1}}T^{1}_{i_{2}},~~\vert 
3>=e_{1}T^{1}_{i_{1}}T^{2}_{i_{2}}=\delta_{i_{1},i_{2}}
\sum_{i}{}^{(s)}T^{1}_{i}T^{2}_{i}.\eqno(12)$$

\vskip .3cm
   The left coset decomposition of $D_{f}(n)$ with respect to the subalgebra
$S_{f_{1}}\times S_{f_{2}}$ is denoted by

$$D_{f}(n)=\sum_{\omega k}\oplus Q^{k}_{\omega}(S_{f_{1}}\times
S_{f_{2}}),\eqno(13)$$

\noindent where the left coset representatives $\{ Q^{k}_{\omega}\}$ 
have two types of operations. One is the order-preserving permutations,

$$Q^{k=0}_{\omega}(\omega^{0}_{1},~\omega^{0}_{2})=(\omega_{1},~\omega_{2}),
\eqno(14)$$

\noindent where

$$(\omega_{1})=(a_{1},~a_{2},~\cdots,~a_{f_{1}}),~~(\omega_{2})
=(a_{f_{1}+1},~a_{f_{1}+2},~\cdots,~a_{f})\eqno(15)$$

\noindent with $a_{1}<a_{2}<\cdots <a_{f_{1}}$,
$a_{f_{1}+1}<a_{f_{1}+2}<\cdots <a_{f}$, and $a_{i}$ represents any
one of the numbers $1,~2,~\cdots,~f$. The other, $\{ Q^{k\geq 1}_{\omega}\}$
contains $k$-time trace contractions between two sets of indices $(\omega_{1})$
and $(\omega_{2})$. For example, in $S_{2}\times S_{1}\uparrow D_{3}(n)$
for the outer product $[2]\times [1]$, there are six elements in
$\{ Q^{k}_{\omega}\}$ with 

$$\{ Q^{0}_{\omega}\}=\{ 1,~g_{2},~g_{1}g_{2}\},~~\{Q^{1}_{\omega}\}
=\{ e_{2},~g_{1}e_{2},~e_{1}g_{2}\}.\eqno(16)$$

\vskip .3cm
 The ordering of the sequences $(\omega )$ is specified in the following way. If
there is no trace contraction, we regard the part $(\omega_{1})=(a_{1},~a_{2},~\cdots,
a_{f_{1}})$ as a vector of length $f_{1}$. If the last nonzero component of the
vector $(\omega_{1})-(\bar{\omega}_{1})$ is less than zero, then we say $(\omega )
\leq (\bar{\omega_{1}})$. This ordering of $(\omega_{1},~\omega_{2})$ is consistent
with that for symmetric groups [chen]. If there is $k$-time trace contraction, we regard
$\omega^{k}$ as vector of length $k$ with the components $(a_{i_{1}}a_{i_{1}^{\prime}})
(a_{i_{2}}a_{i_{2}^{\prime}})\cdots (a_{i_{k}}a_{i_{k}^{\prime}})$. If the last nonzero component
of the vector $\omega^{k}-\bar{\omega}^{k}$ is less than zero, we say $\omega^{k}<\bar{\omega}^{k}$.
The total order of $(\overbrace{\omega_{1})~(\omega_{2}}^{k})$ is specified by
$k=0,~1,~2,~\cdots,~\min(f_{1},f_{2})$, where $(\overbrace{\omega_{1}),~(\omega_{2}}^{k})$ stands for 
$k$-time contractions between indices in $(\omega_{1})$ and $(\omega_{2})$.
For example, in $S_{2}\times S_{1}\uparrow D_{3}(n)$ for
the outer-product $[2]\times [1]$, the six elements are arranged as
$\{1,~g_{2},~g_{1}g_{2},~e_{1}g_{2},~g_{1}e_{2},~e_{2}\}$.
\vskip .3cm
 The uncoupled basis vectors needed in construction of
the coupled basis vectors of $[\lambda ]$ for $D_{f}(n)$, are denoted by

$$Q^{k}_{\omega}\vert Y^{[\lambda_{1}]}_{m_{1}},~Y_{m_{2}}^{[\lambda_{2}]},~(
\omega^{0}_{1}),~(\omega_{2}^{0})>=\vert
Y^{[\lambda_{1}]}_{m_{1}},~Y^{[\lambda_{2}]}_{m_{2}},~(\overbrace{\omega_{1}),~
(\omega_{2}}^{k})>,\eqno(17)$$

\noindent 
The basis vectors of $[\lambda ]_{f-2k}$ can thus be expressed in terms of the uncoupled
basis vectors given by (17):

$$\vert [\lambda ]_{f-2k},~\tau;~\rho >=
\sum_{m_{1}~m_{2}~\omega~k^{\prime}}C^{[\lambda]_{f-2k}~\rho;\tau}_{m_{1}m_{2};k^{\prime}\omega}
Q^{k^{\prime}}_{\omega}\vert Y^{[\lambda_{1}]}_{m_{1}}(\omega_{1}^{0}),~Y^{[\lambda_{2}]}
_{m_{2}}(\omega^{0}_{2})>,\eqno(18)$$

\noindent where $\rho$ is the multiplicity label needed in the outer-product
$[\lambda_{1}]_{f_{1}}\times [\lambda_{2}]_{f_{2}}\uparrow[\lambda]_{f-2k}$,
$\tau$ stands for other labels needed for the irrep $[\lambda ]_{f-2k}$,
$0\leq k^{\prime}\leq k$, and
the coefficient $C^{[\lambda ]_{f-2k}~\rho;\tau}_{m_{1}m_{2};k^{\prime}\omega}$ is
$[\lambda_{1}]_{f_{1}}\times[\lambda_{2}]_{f_{2}}\uparrow[\lambda]_{f-2k}$
Induction coefficient (IDC) or the Outer-product reduction coefficient (ORC).
\vskip .3cm
   The IDCs satisfy the following orthogonality relation:

$$\sum_{m_{1}m_{2}k^{\prime}\omega m^{\prime}_{1}m^{\prime}_{2}k^{\prime\prime}\omega^{\prime}}
C^{[\lambda ]_{f-2k}~\rho;\tau}_{m_{1}m_{2};k^{\prime}\omega}
C^{[\lambda^{\prime}]_{f-2k}~\rho^{\prime};\tau^{\prime}}_{m^{\prime}_{1}m^{\prime}_{2};k^{\prime\prime}\omega^{\prime}}
{\cal N}^{[\lambda_{1}\lambda_{2}]}_{m_{1}m_{2}k^{\prime}\omega; m^{\prime}_{1}m^{\prime}_{2}k^{\prime\prime}\omega^{\prime}}=
\delta_{\lambda\lambda^{\prime}}\delta_{\tau\tau^{\prime}}\delta_{\rho\rho^{\prime}},\eqno(19)$$

\noindent where  ${\cal N}^{[\lambda_{1}][\lambda_{2}]}$ is symmetric norm matrix, of which the elements are
defined $^{[8]}$ by

$${\cal N}^{[\lambda_{1}][\lambda_{2}]}_{m_{1}m_{2}k^{\prime}\omega;
m^{\prime}_{1}m^{\prime}_{2}k^{\prime\prime}\omega^{\prime}}=
<Y^{[\lambda_{1}]}_{m_{1}}Y^{[\lambda_{2}]}_{m_{2}};(\omega_{1}^{0})(\omega_{2}^{0})\vert
Q^{k^{\prime}}_{\omega}Q^{k^{\prime\prime}}_{\omega^{\prime}}\vert
Y^{[\lambda_{1}]}_{m^{\prime}_{1}}Y^{[\lambda_{2}]}_{m^{\prime}_{2}};(\omega_{1}^{0})(\omega_{2}^{0})>.\eqno(20)$$

\noindent These matrix elements can easily be calculated by using the algebraic relations of Brauer
algebra given by (1) and (2) and those given in [8]. While the coupled basis vectors  $\vert [\lambda ]_{f-2k},
\tau; \rho>$ are orthonormal.

$$< [\lambda^{\prime}]_{f-2k^{\prime}};\tau^{\prime},\rho^{\prime}\vert  [\lambda]_{f-2k};\tau,\rho>=
\delta_{\lambda\lambda^{\prime}}\delta_{\tau\tau^{\prime}}\delta_{\rho\rho^{\prime}}\delta_{kk^{\prime}}.\eqno(21)$$
\vskip 2cm
\begin{centering}
{\large 3. Evaluation of the IDCs}\\
\end{centering}
\vskip .3cm
   The linear equation method (LEM) has been proved effective in deriving
SDCs and IDCs of Hecke algebra,$^{[14]}$ as well as  SDCs of Brauer algebra.$^{[10]}$
The procedure for the evaluation of the IDCs of $S_{f_{1}}\times  S_{f_{2}}\uparrow
D_{f}(n)$  is similar to that proposed in [8]. 
\vskip .3cm
   Firstly, applying the operators $R_{i}~(=g_{i}~{\rm or}~e_{i})$ with $i=1,2,\cdots,
f_{1}+f_{2}-1$ to (18), the left-hand side of (18) becomes

$$\sum_{~~m_{1}m_{2}\omega k^{\prime}}\sum_{\rho^{\prime}\tau^{\prime}}
C^{[\lambda ]_{f-2k}~\rho^{\prime};\tau^{\prime}}_{m_{1}m_{2};k^{\prime}\omega~}
<[\lambda ]_{f-2k}\rho^{\prime}\tau^{\prime}\vert R_{i}\vert [\lambda
]_{f-2k}\rho\tau>Q^{k^{\prime}}_{\omega}\vert Y^{[\lambda_{1}]}_{m_{1}}Y^{[\lambda_{2}]}_{m_{2}},
(\omega^{0}_{1})(\omega_{2}^{0})>.\eqno(22)$$
\vskip .3cm
\noindent While the right-hand side of (18) becomes

$$\sum_{m_{1}m_{2}\omega k^{\prime}}C^{[\lambda]_{f-2k}\rho\tau}_{m_{1}m_{2},k^{\prime}\omega}
(R_{i}Q^{k^{\prime}}_{\omega})\vert Y^{[\lambda_{1}]}_{m_{1}}Y^{[\lambda_{2}]}_{m_{2}},
(\omega_{1}^{0})(\omega_{2}^{0})>.\eqno(23)$$

\noindent Then, combining (22) and (23), we get

$$\sum_{\rho^{\prime}\tau^{\prime}}C^{[\lambda]_{f-2k}\rho^{\prime}\tau^{\prime}}_{m_{1}m_{2},k^{\prime}\omega}
<[\lambda ]_{f-2k}\rho^{\prime}\tau^{\prime}\vert R_{i}\vert [\lambda
]_{f-2k}\rho\tau>=C^{[\lambda]_{f-2k}\rho^{\prime}\tau^{\prime}}_{m_{1}m_{2},k^{\prime}\omega^{\prime}}f_{i},
\eqno(24)$$

\noindent where $C^{[\lambda]_{f-2k}\rho^{\prime}\tau^{\prime}}_{m_{1}m_{2},k^{\prime}\omega^{\prime}}f_{i}$
is the coefficient in front of $Q^{k^{\prime}}_{\omega}\vert Y^{[\lambda_{1}]}_{m_{1}}Y^{[\lambda_{2}]}_{m_{2}},
(\omega_{1}^{0})(\omega_{2}^{0})>$ after applying $R_{i}$ to the right-hand side of (18), and  
$<[\lambda ]_{f-2k}\rho^{\prime}\tau^{\prime}\vert R_{i}\vert [\lambda ]_{f-2k}\rho\tau>$ is
matrix elements of $R_{i}$ in the standard basis given in (4), which have already been known.$^{[8]}$ 
\vskip .3cm
   The linear relations or so-called  a part of the intertwining relations among the IDCs
given by  (24) are  sufficient to determine these IDCs up to a normalization factor,$^{[8]}$ which
can then be calculated by using the orthogonality relation (19).  It will be shown that
the CGCs of SO(n), expressed in terms of these IDCs, need to be normalized again  according to different
cases. Therefore, the normalization  of these IDCs is not necessary. However, the sign of the
normalization factors, which gives overall phase of the IDCs should be chosen
beforehand. In our calculation, the overall phase is fixed by requiring that the
IDCs with $\min(\tau)$ at first, then with $\min(m_{1})$, and smallest indices $\omega$ and $k^{\prime}$
be positive

$$C^{[\lambda]_{f-2k}\rho\min(\tau)}_{\min(m_{1})m_{2},k^{\prime}=0\min(\omega)}>~0\eqno(25)$$
\vskip .3cm 
Using the algebraic relations
of Brauer algebra, Eq. (24), and irreducible representations of symmetric 
groups in the standard
basis,$^{[15]}$ one can obtain all the IDCs of 
$S_{f_{1}}\times S_{f_{2}}\uparrow D_{f}$. In what 
follows, we will give a simple example of deriving the IDCs and some basic 
features of these coefficients.

\vskip .3cm
\noindent {\bf Example 1.} Deriving  IDCs of $S_{1}\times S_{1}\uparrow D_{2}(n)$. The outer product
reduction is $[1]\times [1]\uparrow [2]+[1^{2}]+[0]$. In this case, Eq. (18) can be written as

$$\vert [2]>=\sum_{i=1}^{3}a_{i}\vert i>,~\vert [1^{2}]>=\sum_{i=1}^{3}b_{i}\vert i>,
~~\vert [0]>=\sum_{i=1}^{3}c_{i}\vert i>,\eqno(26a)$$

\noindent where $a_{i}$, $b_{i}$, and $c_{i}$ are the corresponding IDCs, and $\vert i>$ $(i=1,2,3)$
are the uncoupled basis vectors defined by

$$\vert 1>=\vert 1,2>,~~\vert 2>=g_{1}\vert 1>,~~\vert 3>=e_{1}\vert 1>.\eqno(26b)$$

\noindent  Applying generators $g_{1}$ and $e_{1}$, respectively, to (26a), one obtains

$$a_{1}=a_{2},~~a_{3}=-{2\over{n}}a_{1},$$

$$b_{1}=-b_{2},~~b_{3}=0,$$

$$c_{1}=c_{2}=0,~~c_{3}\neq 0.\eqno(26c)$$

\noindent The norm matrix for this case is

$${\cal N}^{[1][1]}=\left(\matrix{
1 &\delta_{i_{1}i_{2}} &\delta_{i_{1}i_{2}}\cr
\delta_{i_{1}i_{2}} &1 &\delta_{i_{1}i_{2}}\cr
\delta_{i_{1}i_{2}} &\delta_{i_{1} i_{2}} &n\delta_{i_{1}i_{2}}\cr}\right),
\eqno(26d)$$

\noindent which can be proved by using (8)-(10).
 Hence, the coupled basis vectors can now be written as

$$\vert [2]>=a_{1}\left(\vert 1>+\vert 2>-{2\over{n}}\vert 3>\right),$$

$$\vert [1^{2}]>=b_{1}\left(\vert 1>-\vert 2>\right),$$

$$\vert [0]>=c_{3}\vert 3>.\eqno(27)$$

\noindent Using the norm matrix (26d), one can check that  basis vectors given by (27) are
orthogonal. The normalization factors, of which the signs should be chosen according to (25),
 can easily be obtained  by using (21) and (26).

$$a_{1}=\sqrt{n\over{2(n+\delta_{i_{1}i_{2}}(n-2))}},~~ b_{1}=\sqrt{1\over{2}},
~~c_{3}=\sqrt{1\over{n}}.\eqno(28)$$

  It can easily be seen that $\vert 3>$ is a null vector when $i_{1}\neq i_{2}$. In this case,
(27) becomes outer-product basis vectors of symmetric group  $S_{1}\times S_{1}\uparrow S_{2}$.
 It is clear that the induced representations of $D_{f}(n)$ from $S_{f_{1}}\times S_{f_{2}}$
are $SO(n)$ tensor component dependent. Actually, The normalization of these basis vectors
is not necessary with respect to representations of $D_{f}(n)$.
On the other hand, It can be easily seen from (26)-(28) that
normalization factors of the IDC's  are also $SO(n)$ tensor component dependent. 
The situation will become more complicated when $f_{1}+f_{2}=f\geq 3$.
Furthermore,
our purpose is to evaluate CGCs of $O(n)$ from these IDCs. The coupled basis vectors of 
$O(n)$ obtained from these IDCs through assimilation
need to be normalized again. Therefore,  we only list
unnormalized IDCs of $S_{f_{1}}\times S_{f_{2}}\uparrow D_{f}(n)$. 
The method of how to evaluate $O(n)$ CGCs from these unnormalized 
IDCs will be presented in the next section.
\vskip .3cm
   	In the following, we list unnormalized IDCs of $S_{f_{1}}\times S_{f_{2}}\uparrow D_{f}(n)$
with $f_{1}+f_{2}=f\leq 4$.  The signs of the normalization factors given below are all
chosen to be positive, which is fixed by our phase convention (25). Only the absolute values
of these normalization factors need to be determined according to different $SO(n)$ tensor
components later.
\vskip .7cm

\noindent (1) $D_{1}(n)\times D_{1}(n)\uparrow D_{2}(n)$ ~for~ $[1]\times [1]= [2]+[1^{2}]+[0].$
\vskip .3cm
\begin{tabbing}
11\=2222222222222222222222222222222222222222222222222222222222222222222222222222222222\=\kill\\
\>{$\left\vert [2]\right>=a_{1}\left(\vert 1>+\vert 2>-{2\over{n}}\vert 3>\right),$}\\
\>{}\\
\>{$\left\vert [1^{2}]\right>=\sqrt{1\over{2}}\left( \vert 1>-\vert 2>\right),$}\\
\>{}\\
\>{$\left\vert [0]\right>=\sqrt{1\over{n}}\vert 3>,$}\\
\end{tabbing}
\noindent where  $\vert 1>=\vert 1,~2>$, $\vert 2>=g_{1}\vert 1>$,
$\vert 3>=e_{1}\vert 1>$.
\vskip 1cm
\noindent (2) $D_{2}(n)\times D_{1}(n)\uparrow D_{3}(n)$~ for~ $[2]\times [1]=[3]+[21]+[1].$
\vskip .3cm
\begin{tabbing}
11\=2222222222222222222222222222222222222222222222222222222222222222222222222222222222\=\kill\\
\>{$\left\vert [3]\right>=a_{1}\left(\vert 1>+\vert 2>+\vert 3>
-{2\over{n+2}}(\vert 4>+\vert 5>+\vert 6>)\right),$}\\
\>{}\\
\>{$\left\vert [21]_{1}\right>={1\over{\sqrt{3}}}a_{2}\left(
2\vert 1>-\vert 2>-\vert 3>+{1\over{n-1}}(2\vert 4>-\vert 5>-\vert 6>)\right),$}\\
\>{}\\
\>{$\left\vert [21]_{2}\right>=a_{2}\left(\vert 2>-\vert 3>+{1\over{n-1}}(\vert 5>-\vert 6>)\right),$}\\
\>{}\\
\>{$\left\vert [1][0]\right>=a_{3}\vert 4>,$}\\
\>{}\\
\>{$\left\vert [1][2]\right>=\sqrt{1\over{2(n+2)(n-1)}}a_{3}\left(2\vert 4>-n(\vert 5>+\vert
6>)\right),$}\\
\>{}\\
\>{$\left\vert [1][1^{2}]\right>=\sqrt{n\over{2(n-1)}}a_{3}\left(\vert 5>-\vert 6>\right),$}\\
\end{tabbing}
\noindent where  $\vert 1>=\vert 12,3>$, $\vert 2>=g_{2}\vert 1>$,
$\vert 3>=g_{1}g_{2}\vert 1>$, $\vert 4>=e_{1}g_{2}\vert 1>$,
$\vert 5>=g_{1}e_{2}\vert 1>$, $\vert 6>=e_{2}\vert 1>$.
\vskip 1cm
\noindent (3) $D_{2}(n)\times D_{1}(n)\uparrow D_{3}(n)$~ for~ $[1^{2}]\times [1]=[1^{3}]+[21]+[1].$
\begin{tabbing}
11\=2222222222222222222222222222222222222222222222222222222222222222222222222222222222\=\kill \\
\>{$\left\vert [1^{3}]\right>={1\over{\sqrt{3}}}\left(\vert 1>-\vert 2>+\vert 3>\right),$}\\
\>{}\\
\>{$\left\vert [21]_{1}\right>=\sqrt{3}a_{1}\left(\vert 2>+\vert 3>-{1\over{n-1}}
(\vert 4>+\vert 5>+\vert 6>)\right),$}\\
\>{}\\
\>{$\left\vert [21]_{2}\right>=a_{1}\left(2\vert 1>+\vert 2>-\vert 3>+{3\over{n-1}}(\vert 5>-\vert 6>)\right),$}\\
\>{}\\
\>{$\left\vert [1][0]\right>=a_{2}\vert 4>,$}\\
\>{}\\
\>{$\left\vert [1][2]\right>={1\over{\sqrt{2(n+2)(n-1)}}}a_{2}\left(2\vert 4>+n(\vert 5>+\vert 6>)\right),$}\\
\>{}\\
\>{$\left\vert [1][1^{2}]\right>=\sqrt{n\over{2(n-1)}}a_{2}\left(\vert 6>-\vert 5>\right),$}\\
\end{tabbing}
\noindent where  
$\vert 1>=\vert
\begin{array}{l}
1\\
2
\end{array},3>$, 
$\vert 2>=g_{2}\vert 1>$,
$\vert 3>=g_{1}g_{2}\vert 1>$, $\vert 4>=e_{1}g_{2}\vert 1>$,
$\vert 5>=g_{1}e_{2}\vert 1>$, $\vert 6>=e_{2}\vert 1>$.
\vskip 1cm
\noindent (4) $D_{3}(n)\times D_{1}(n)\uparrow D_{4}(n)$~ for~ $[3]\times [1]=[4]+
[31]+[2].$
\begin{tabbing}
11\=2222222222222222222222222222222222222222222222222222222222222222222222222222222222\=\kill\\
\>{$\left\vert [4]\right>=c_{1}\left(\vert 1>+\vert 2>+\vert 3>+\vert 4>
-{2\over{n+4}}(\vert 5>+\vert 6>+\vert 7>+\vert 8>+\vert 9>+\vert 10>)\right),$}\\
\>{}\\
\>{$\left\vert [31]_{1}\right>={c_{2}\over{\sqrt{2}}}\left(
3\vert 1>-\vert 2>-\vert 3>-\vert 4>+
{2\over{n}}(\vert 5>+\vert 6>-\vert 7>+\vert 8>-\vert 9>-\vert 10>)\right),$}\\
\>{}\\
\>{$\left\vert [31]_{2}\right>=c_{2}\left( 2\vert 2>-\vert 3>-\vert 4>+
{1\over{n}}(2\vert 5>-\vert 6>+\vert 7>-\vert 8>+\vert 9>-2\vert 10>)\right),$}\\
\>{}\\
\>{$\left\vert [31]_{3}\right>=\sqrt{3}c_{2}\left(\vert 3>-\vert 4>+{1\over{n}}(
\vert 6>+\vert 7>-\vert 8>-\vert 9>)\right),$}\\
\>{}\\
\>{$\left\vert [2][1][0]\right>=c_{3}\vert 5>,$}\\
\>{}\\
\>{$\left\vert [2][1][2]\right>={1\over{\sqrt{2(n+2)(n-1)}}}c_{3}\left(
2\vert 5>-n(\vert 6>+\vert 8>)\right),$}\\
\>{}\\
\>{$\left\vert [2][1][1^{2}]\right>=\sqrt{2\over{(n+2)(n-1)}}c_{3}\left(\vert 6>-\vert 8>\right),$}\\
\>{}\\
\>{$\left\vert [2][3]\right>=\sqrt{4\over{3n(n+4)}}c_{3}\left({2\over{n+2}}
(\vert 5>+\vert 6>+\vert 8>)-\vert 7>-\vert 9>-\vert 10>)\right),$}\\
\>{}\\
\>{$\left\vert [21]_{1}\right>=\sqrt{2(n-1)\over{3(n^{2}-4)}}c_{3}
\left(2\vert 10>-\vert 7>-\vert 9>+{1\over{n+1}}(2\vert 5>-\vert 6>-\vert 8>)\right),$}\\
\>{$\left\vert [21]_{2}\right>=\sqrt{2(n-1)\over{(n^{2}-4)}}c_{3}
\left(\vert 9>-\vert 10>+{1\over{n+1}}(\vert 6>-\vert 8>)\right),$}\\
\end{tabbing}
\noindent where $\vert 1>=\vert 123,4>$, $\vert 2>=g_{3}\vert 1>$, $\vert 3>=g_{2}g_{3}\vert 1>$,
$\vert 4>=g_{1}g_{2}g_{3}\vert 1>$,
$\vert 5>=e_{1}g_{2}g_{3}\vert 1>$,
$\vert 6>=g_{1}e_{2}g_{3}\vert 1>$, $\vert 7>= g_{1}g_{2}e_{3}\vert 1>$,
$\vert 8>=e_{2}g_{3}\vert 1>$,
$\vert 9>=g_{2}e_{3}\vert 1>$, $\vert 10>=e_{3}\vert 1>$.
\vskip .5cm
\noindent (5) $D_{3}(n)\times D_{1}(n)\uparrow D_{4}(n)$~ for~ $[1^{3}]\times [1]=
[1^{4}]+[211]+[1^{2}].$
\vskip .3cm
\begin{tabbing}
11\=2222222222222222222222222222222222222222222222222222222222222222222222222222222222\=\kill\\
\>{$\left\vert [1^{4}]\right>={1\over{2}}\left( \vert 1>-\vert 2>+\vert 3>-\vert 4>\right)$,}\\
\>{}\\
\>{$\left\vert [21^{2}]_{1}\right>={(n-2)\over{\sqrt{2}}}c_{4}\left(
3\vert 1>+\vert 2>-\vert 3>+\vert 4>-{4\over{n-2}}(\vert 7>-\vert 9>+\vert 10>)\right),$}\\
\>{}\\
\>{$\left\vert [21^{2}]_{2}\right>=(n-2)c_{4}\left( 2\vert 2>+\vert 3>-\vert 4>+{1\over{n-2}}
(3\vert 6>+\vert 7>-3\vert 8>-\vert 9>-2\vert 10>)\right)$,}\\
\>{}\\
\>{$\left\vert [21^{2}]_{3}\right>=\sqrt{3}(n-2)c_{4}\left(\vert 3>+\vert 4>-
{1\over{n-2}}(2\vert 5>+\vert 6>+\vert 7>+\vert 8>+\vert 9>)\right),$}\\
\>{}\\
\>{$\left\vert [1^{2}][1][0]\right>=c_{5}\vert 5>,$}\\
\>{}\\
\>{$\left\vert [1^{2}][1][2]\right>={1\over{\sqrt{2(n+2)(n-1)}}}c_{5}\left(
2\vert 5>+n(\vert 6>+\vert 8>)\right)$,}\\
\>{}\\
\>{$\left\vert [1^{2}][1][1^{2}]\right>=\sqrt{n\over{2(n-1)}}c_{5}\left(\vert 8>-\vert 6>\right),$}\\
\>{}\\
\>{$\left\vert [1^{2}][21]_{1}\right>=\sqrt{n(n-1)\over{2(n^{2}-4)}}c_{5}
\left(\vert 7>+\vert 9>+{1\over{n-1}}(2\vert 5>+\vert 6>+\vert 8>)\right),$}\\
\>{}\\
\>{$\left\vert [1^{2}][21]_{2}\right>=-\sqrt{n(n-1)\over{6(n^{2}-4)}}c_{5}\left(
{3\over{n-1}}(\vert 6>-\vert 8>)+\vert 7>-\vert 9>-2\vert 10>\right),$}\\
\>{}\\
\>{ $\left\vert [1^{2}][1^{3}]\right>={1\over{3}}\sqrt{n\over{n-2}}c_{5}\left(\vert 7>-\vert
9>+\vert 10>\right),$}\\
\end{tabbing}
\noindent  where $\vert 1>=\left\vert
\begin{array}{l}
1\\
2\\
3
\end{array},4\right>$, $\vert 2>=g_{3}\vert 1>$, $\vert 3>=g_{2}g_{3}\vert 1>$, 
$\vert 4>=g_{1}g_{2}g_{3}\vert 1>$,
$\vert 5>=e_{1}g_{2}g_{3}\vert 1>$, 
$\vert 6>=g_{1}e_{2}g_{3}\vert 1>$, $\vert 7>= g_{1}g_{2}e_{3}\vert 1>$,
$\vert 8>=e_{2}g_{3}\vert 1>$,
$\vert 9>=g_{2}e_{3}\vert 1>$, $\vert 10>=e_{3}\vert  1>$.
\vskip 1cm
\noindent (6) $D_{3}(n)\times D_{1}(n)\uparrow D_{4}(n)$~ for~ $[21]\times [1]= [31]+[22]+[211]+[1^{2}]+[2]$.
\vskip .3cm
\begin{tabbing}
333\=11111111\=2222222222222222222222222222222222222222222222222222222222222222222222222222222222\=\kill\\
\>{$\left\vert [31]_{1}\right>=$}\>{${n(n+2)\over{\sqrt{3}}}d_{1}
\left(\vert 2>+\vert 3>+\vert 4>
+{1\over{2n}}(\vert 13>+\vert 18>-2\vert 19>+\sqrt{3}(\vert 14>-\vert 17>))
\right.$}\\
\>{}\>{$\left.-{(2n+1)\over{n(n+2)}}(\vert 9>+\vert 11>+\vert 16>)-{\sqrt{3}
\over{n(n+2)}}(\vert 10>+\vert 12>
-\vert 15>)\right),$}\\
\>{}\\
\>{$\left\vert [31]_{2}\right>=$}\>{${\sqrt{6}n(n+2)\over{4}}d_{1}\left(
\vert 1>+{1\over{3}}\vert 2>-{1\over{6}}(\vert 3>+\vert 4>)+{\sqrt{3}\over{2}}(\vert 7>+\vert 8>)
+\right.$}\\
\>{}\>{${(n-4)\over{6n(n+2)}}(2\vert 9>-\vert 11>-\vert 16>)
+{\sqrt{3}(3n+4)\over{6n(n+2)}}(\vert 12>-\vert 15>-2\vert 10>)$}\\
\>{}\>{$\left.+{\sqrt{3}(n-4)\over{6n(n+2)}}(\vert 14>-\vert 17>)-
{(11n+4)\over{6n(n+2)}}
(\vert 13>+\vert 18>)-{4\over{3n}}\vert 19>\right)$}\\
\>{}\\
\>{$\left\vert [31]_{3}\right>=$}\>{${\sqrt{2}n(n+2)\over{8}}d_{1}\left(
\vert 3>-\vert 4>+\sqrt{3}(\vert 7>-\vert 8>)+2\sqrt{3}(\vert 5>+\vert 6>)+{(n-4)\over{n(n+2)}}(\vert 13>\right.$}\\
\>{}\>{$\left.+\vert 11>-\vert 16>-\vert 18>)
-{\sqrt{3}(3n+4)\over{n(n+2)}}(\vert 12>+\vert 15>+\vert 14>+\vert 17>)-{4\sqrt{3}\over{n+2}}\vert 20>\right)$,}\\
\\
\>{$\left\vert [21^{2}]_{1}\right>=$}\>{${3\sqrt{2}\over{4}}d_{2}\left(\vert 1>-\vert 2>+
{\sqrt{3}\over{6}}(\vert 7>+\vert 8>-\sqrt{3}\vert 3>+\sqrt{3}\vert 4>)+{\sqrt{3}\over{6(n-2)}}(
2\vert 10>-2\sqrt{3}\vert 9>\right.$}\\
\>{}\>{$\left.-\vert 12>+\vert 15>+\vert 14>-\vert 17>)+{1\over{2(n-2)}}(\vert 11>+\vert 16>-
\vert 13>-\vert 18>)\right)$,}\\
\>{}\\
\>{$\left\vert [21^{2}]_{2}\right>={3\sqrt{3}\over{4}}d_{2}\left(\vert 4>-\vert 3>+
{\sqrt{3}\over{9}}(6\vert 5>-2\vert 6>-\vert 7>+\vert 8>)+{1\over{9(n-2)}}(3\vert 13>-3\vert 18>\right.$}\\
\>{}\>{$\left.+3\sqrt{3}(\vert 12>+\vert 15>)-9\vert 11>+9\vert 16>-\sqrt{3}\vert 14>
-\sqrt{3}\vert 17>-4\sqrt{3}\vert 20>)\right),$}\\
\>{}\\
\>{$\left\vert [21^{2}]_{3}\right>=$}\>{$d_{2}\left(\vert 6>-\vert 7>+\vert 8>
+{1\over{2(n-2)}}(\vert 14>+\vert 17>-\sqrt{3}(\vert 13>-\vert 18>)-2\vert 20>)\right)$,}\\
\>{}\\
\>{$\left\vert [22]_{1}\right>=$}
\>{$d_{3}\left(\vert 5>-\vert 6>+{1\over{n-2}}(\vert 12>+\vert 15>-\vert 14>-\vert 17>)\right)$,}\\
\>{}\\
\>{$\left\vert [22]_{2}\right>=$}
\>{$d_{3}\left(\vert 1>-{1\over{n-2}}\vert 19>)\right)$,}\\
333\=1111111111\=2222222222222222222222222222222222222222222222222222222222222222222222222222222222\=\kill\\
\>{$\left\vert [2][1][0]\right>=$}
\>{$d_{4}\left(\vert 9>+\sqrt{3}\vert 10>)\right)$,}\\
\>{}\\
\>{$\left\vert [2][1][2]\right>=$}
\>{$\sqrt{2\over{(n+2)(n-1)}}d_{4}\left(\vert 9>+\sqrt{3}\vert 10>
-{n\over{2}}(\vert 11>+\vert 16>)-{\sqrt{3}n\over{2}}(\vert 12>-\vert 15>)\right)$,}\\
\>{}\\
\>{$\left\vert [2][1][1^{2}]\right>=$}
\>{$\sqrt{n\over{2(n-1)}}d_{4}\left(\vert 11>-\vert 16>)
+\sqrt{3}(\vert 12>+\vert 15>)\right)$,}\\
\>{}\\
333\=111111111\=2222222222222222222222222222222222222222222222222222222222222222222222222222222222\=\kill\\
\>{$\left\vert [2][3]\right>=$}
\>{${2\over{\sqrt{3(n+2)(n+4)}}}d_{4}\left(\vert 9>+\vert 11>+\vert 16>
+\sqrt{3}(\vert 10>+\vert 12>-\vert 15>)\right.$}\\
\>{}\>{$\left.-{n+2\over{2}}(\vert 13>+\vert 18>-2\vert 19>+\sqrt{3}\vert 14>
-\sqrt{3}\vert 17>)\right),$}\\
\>{}\\
\>{$\left\vert [2][21]_{1}\right>=$}
\>{$\sqrt{2\over{3(n-2)(n-1)}}d_{4}\left(\vert 9>+\sqrt{3}\vert 10>-{1\over{2}}(\vert 11>+\vert 16>)
-{\sqrt{3}\over{2}}(\vert 12>-\vert 15>)\right.$}\\
\>{}\>{$\left.-{n-1\over{2}}(\vert 13>+\vert 18>)-{n-1\over{2}}\sqrt{3}(\vert 14>
-\vert 17>)-2(n-1)\vert 19>\right),$}\\
\>{}\\
\>{$\left\vert [2][21]_{2}\right>=$}
\>{$\sqrt{1\over{2(n-2)(n-1)}}d_{4}\left(\vert 11>-\vert 16>+\sqrt{3}(\vert 12>+
\vert 15>)
-(n-1)(\vert 13>-\vert 18>)\right.$}\\
\>{}\>{$\left.-(n-1)\sqrt{3}(\vert 14>
+\vert 17>)\right),$}\\
333\=111111111111\=2222222222222222222222222222222222222222222222222222222222222222222222222222222222\=\kill\\
\>{$\left\vert [1^{2}][1][0]\right>=$}
\>{$d_{5}\left(\vert 9>-\sqrt{3}\vert 10>)\right)$,}\\
\>{}\\
\>{$\left\vert [1^{2}][1][2]\right>=$}
\>{$\sqrt{2\over{(n+2)(n-1)}}d_{5}\left(\vert 9>-\sqrt{3}\vert 10>
-{n\over{2}}(\vert 11>+\vert 16>)-{\sqrt{3}n\over{2}}(\vert 12>-\vert 15>)\right)$,}\\
\>{}\\
\>{$\left\vert [1^{2}][1][1^{2}]\right>=$}
\>{$\sqrt{n\over{2(n-1)}}d_{5}\left(\vert 11>-\vert 16>)
-\sqrt{3}(\vert 12>+\vert 15>)\right)$,}\\
\>{}\\
\>{$\left\vert [1^{2}][21]_{1}\right>=$}
\>{$\sqrt{2n\over{(n^{2}-4)(n-1)}}d_{5}\left(\vert 9>-{\sqrt{3}\over{n}}\vert 10>
-{1\over{2}}(\vert 11>+\vert 16>)
+{\sqrt{3}\over{2}}(\vert 12>-\vert 15>)\right.$}\\
\>{}\>{$\left.+{n-1\over{2}}(\vert 13>+\vert 18>)-{n-1\over{2}}\sqrt{3}(\vert 14>
-\vert 17>)\right),$}\\
\>{}\\
\>{$\left\vert [1^{2}][21]_{2}\right>=$}
\>{$\sqrt{3\over{2(n^{2}-4)(n-1)}}d_{5}\left(\vert 11>-\vert 16>+{1\over{\sqrt{12n}}}
(\vert 12>+
\vert 15>)
-(n-1)\sqrt{n}(\vert 13>-\vert 18>)\right.$}\\
\>{}\>{$\left.-(n-1)\sqrt{n\over{3}}(\vert 14>
+\vert 17>)+(n-1){\sqrt{2n\over{3}}}\vert 20>\right),$}\\
\>{}\\
\>{$\left\vert [1^{2}][1^{3}]\right>=$}
\>{$\sqrt{n\over{2(n-2)}}d_{5}\left(\sqrt{3}\vert 13>-\vert 14>-\vert 17>
-\sqrt{3}\vert 18>+2\vert 20>\right)$,}\\
\end{tabbing}
\noindent where  $\vert 1>=\left\vert 
\begin{array}{l}
12\\
3
\end{array},4\right>$, $\vert 2>=g_{3}\vert 1>$,
$\vert 3>=g_{2}g_{3}\vert 1>$, $\vert 4>=g_{1}g_{2}g_{3}\vert 1>$,\\
$\vert 5>=\left\vert
\begin{array}{l}
13\\
2
\end{array}, 4\right>$,
$\vert 6>=g_{3}\vert 1>$, $\vert 7>=g_{2}g_{3}\vert 5>$,
$\vert 8>=g_{1}g_{2}g_{3}\vert 5>$,
$\vert 9>=e_{1}g_{2}g_{3}\vert 1>$,
$\vert 10>=e_{1}g_{2}g_{3}\vert 5>$, 
$\vert 11>=g_{2}e_{1}g_{2}g_{3}\vert 1>$,
$\vert 12>=g_{2}e_{1}g_{2}g_{3}\vert 5>$, 
$\vert 13>=g_{3}g_{2}e_{1}g_{2}g_{3}\vert 1>$,
$\vert 14>=g_{3}g_{2}e_{1}g_{2}g_{3}\vert 5>$,
$\vert 15>=e_{2}g_{3}\vert 5>$, 
$\vert 16>=e_{2}g_{3}\vert 1>$, 
$\vert 17>= g_{3}e_{2}g_{3}\vert 5>$,
$\vert 18>=g_{3}e_{2}g_{3}\vert 1>$, 
$\vert 19>=e_{3}\vert 1>$, 
$\vert 20>=e_{3}\vert 5>$.
\vskip 1cm
\noindent (7) $D_{2}(n)\times D_{2}(n)\uparrow D_{4}(n)$~ for~ $[2]\times [2]= [4]+[31]+[22]+[2]+[1^{2}]+[0]$.
\vskip .3cm
\begin{tabbing}
333\=11111111\=2222222222222222222222222222222222222222222222222222222222222222222222222222222222\=\kill\\
\>{$\left\vert [4]\right>=$}
\>{$f_{1}\left({n+4\over{2}}(\vert 1>+\vert 2>+\vert 3>+\vert 4>+\vert 5>+\vert 6>)
-\vert 13>-\vert 14>-\vert 17>\right.$}\\
\>{}\>{$-\vert 18>-\vert 15>-\vert 11>-\vert 9>-\vert 12>-\vert 7>-\vert 8>
-\vert 10>-\vert 16>$}\\
\>{}\>{$\left.+{2\over{n+2}}(\vert 21>+\vert 20>+\vert 19>)\right),$}\\
\>{}\\
\>{$\left\vert [31]_{1}\right>=$}
\>{$\sqrt{2\over{3}}f_{2}\left(\vert 1>+\vert 2>+\vert 3>-\vert 4>-\vert 5>-\vert 6>
+{2\over{n+2}}(3\vert 14>+3\vert 10>+3\vert 8>\right.$}\\
\>{}\>{$\left.-\vert 17>-\vert 18>
-\vert 16>-\vert 12>-\vert 11>-\vert 15>-\vert 13>-\vert 9>-\vert 7>\right),$}\\
\>{}\\
\>{$\left\vert [31]_{2}\right>=$}
\>{$\sqrt{1\over{3}}f_{2}\left(2\vert 1>-2\vert 6>-\vert 2>-\vert 3>+\vert 4>+\vert 5>
+{1\over{n+2}}(4\vert 12>+4\vert 7>+4\vert 16>\right.$}\\
\>{}\>{$\left.-2\vert 15>-2\vert 11>
-2\vert 13>-2\vert 9>-2\vert 17>-2\vert 18>)\right),$}\\
\>{}\\
\>{$\left\vert [31]_{3}\right>=$}
\>{$f_{2}\left(\vert 2>-\vert 3>+\vert 4>-\vert 5>
-{2\over{n+2}}(\vert 13>-\vert 9>+\vert 17>\right.$}\\
\>{}\>{$\left.-\vert 18>+\vert 15>
-\vert 11>)\right),$}\\
\>{}\\
\>{$\left\vert [22]_{1}\right>=$}
\>{$f_{3}\left({2(n-2)\over{\sqrt{3}}}(\vert 1>+\vert 6>)
-(n-2)(\vert 2>+\vert 3>+\vert 4>+\vert 5>)\right.$}\\
\>{}\>{$-\vert 13>-\vert 9>-\vert 11>-\vert 15>-\vert 14>-\vert 10>-\vert 12>-\vert 16>$}\\
\>{}\>{$\left.+2(\vert 17>+\vert 18>+\vert 7>+\vert 8>)-{4\over{n-1}}\vert 19>
+{2\over{n-1}}(\vert 21>+\vert 20>)\right)$,}\\
\>{}\\
\>{$\left\vert [22]_{2}\right>=$}
\>{$f_{3}\left(\sqrt{3}(n-2)(\vert 2>-\vert 3>-\vert 4>+\vert 5>)-
\sqrt{3}(\vert 13>-\vert 9>-\vert 15>\right.$}\\
\>{}\>{$\left.+\vert 11>
+\vert 14>-\vert 10>+\vert 12>-\vert 16>)+{2\sqrt{3}\over{n-1}}(\vert 21>-\vert 20>)\right)$,}\\
333\=1111111111\=2222222222222222222222222222222222222222222222222222222222222222222222222222222222\=\kill\\
\>{$\left\vert [2][1][0]\right>=$}
\>{$f_{4}\left(\vert 7>+\vert 8>-{2\over{n}}\vert 19>)\right)$,}\\
\>{}\\
\>{$\left\vert [2][1][2]\right>=$}
\>{$\sqrt{2\over{(n+2)(n-1)}}f_{4}\left(\vert 7>+\vert 8>+\vert 21>+\vert 20>
-{n\over{2}}(\vert 13>+\vert 9>+\vert 14>+\vert 10>-{2\over{n}}\vert 19>)\right)$,}\\
\>{}\\
\>{$\left\vert [2][1][1^{2}]\right>=$}
\>{$\sqrt{n\over{2(n-1)}}f_{4}\left(\vert 9>-\vert 13>-\vert 14>+\vert 10>
+{2\over{n}}(\vert 21>-\vert 20>)\right)$,}\\
\>{}\\
\>{$\left\vert [2][21]_{1}\right>=$}
\>{$\sqrt{2\over{3(n-2)(n-1)}}f_{4}\left(\vert 7>+\vert 8>+\vert 21>+\vert 20>
-2\vert 19>
-{n-1\over{2}}(\vert 15>+\vert 11>\right.$}\\
\>{}\>{$\left.+\vert 12>+\vert 16>)+(n-1)(\vert 17>+\vert 18>)-{1\over{2}}(\vert 13>
+\vert 9>+\vert 14>+\vert 10>)\right),$}\\
\>{}\\
\>{$\left\vert [2][21]_{2}\right>=$}
\>{$\sqrt{2\over{(n-2)(n-1)}}f_{4}\left(\vert 21>-\vert 20>+{n-1\over{2}}(\vert 15>-
\vert 11>-\vert 12>+\vert 16>)\right.$}\\
\>{}\>{$\left.-{1\over{2}}(\vert 13>-\vert 9>+\vert 14>-\vert 10>)\right)$,}\\
333\=111111111\=2222222222222222222222222222222222222222222222222222222222222222222222222222222222\=\kill\\
\>{$\left\vert [2][3]\right>=$}
\>{${2\over{\sqrt{3(n+2)(n+4)}}}f_{4}\left(\vert 13>+\vert 9>+\vert 7>
+\vert 10>+\vert 14>+\vert 8>\right.$}\\
\>{}\>{$\left.-{n+2\over{2}}(\vert 17>+\vert 18>+\vert 16>+\vert 12>
+\vert 11>+\vert 15>)\right),$}\\
\>{}\\
333\=1111111111\=2222222222222222222222222222222222222222222222222222222222222222222222222222222222\=\kill\\
\>{$\left\vert [1^{2}][1][0]\right>=$}
\>{$f_{5}\left(\vert 7>-\vert 8>\right)$,}\\
\>{}\\
\>{$\left\vert [1^{2}][1][2]\right>=$}
\>{~~$\sqrt{2\over{(n+2)(n-1)}}f_{5}\left(\vert 7>-\vert 8>
-{n\over{2}}(\vert 13>+\vert 9>)-\vert 14>-\vert 10>\right)$,}\\
\>{}\\
\>{$\left\vert [1^{2}][1][1^{2}]\right>=$}
\>{~~$\sqrt{n\over{2(n-1)}}f_{5}\left(\vert 9>+\vert 14>
-\vert 10>-\vert 13>\right)$,}\\
\>{}\\
\>{$\left\vert [1^{2}][21]_{1}\right>=$}
\>{$\sqrt{2n\over{(n^{2}-4)(n-1)}}f_{5}\left(\vert 7>-\vert 8>
+{(n-1)(n+2)\over{2n}}(\vert 11>+\vert 15>-\vert 12>-\vert 16>)\right.$}\\
\>{}\>{$\left.-{1\over{2}}(\vert 13>+\vert 9>-\vert 14>
-\vert 10>)\right),$}\\
\>{}\\
\>{$\left\vert [1^{2}][21]_{2}\right>=$}
\>{$~\sqrt{2(n-1)\over{3(n-2)}}f_{5}\left(\vert 17>-\vert 18>+
{1\over{2}}(\vert 15>-\vert 11>+\vert 12>-\vert 16>)
\right.$}\\
\>{}\>{$\left.-\sqrt{3n\over{2(n-1)^{2}(n+2)}}(\vert 13>
-\vert 9>-\vert 14>+\vert 10>)\right),$}\\
\>{}\\
\>{$\left\vert [1^{2}][1^{3}]\right>=$}
\>{$\sqrt{n+2\over{3(n-2)}}f_{5}\left(\vert 17>-\vert 18>-\vert 15>
+\vert 11>-\vert 12>+\vert 16>\right)$,}\\
333\=111111111111\=2222222222222222222222222222222222222222222222222222222222222222222222222222222222\=\kill\\
\>{$\left\vert [0][1][0]\right>=$}\>{$f_{6}\vert 9>$,}\\
\>{}\\
\>{$\left\vert [0][1][2]\right>=$}\>{${-n\over \sqrt{2(n+2)(n-1)}}f_{6}
\left(\vert 20>+\vert 21>+\frac{2(n+2)(n-1)\sqrt{2(n+2)(n-1)}}{n^{2}}
\vert 19>\right)$,}\\
\>{}\\
\>{$\left\vert [0][1][1^{2}]\right>=$}\>{${\sqrt{n\over{2(n-1)}}}f_{6}
\left( \vert 20>-\vert 21>\right)$,}\\
\end{tabbing}
\noindent where  $\vert 1>=\vert 12,34>$, $\vert 2>=g_{2}\vert 1>$,
$\vert 3>=g_{1}g_{2}\vert 1>$, $\vert 4>=g_{3}g_{2}\vert 1>$,
$\vert 5>=g_{1}g_{3}g_{2}\vert 1>$,
$\vert 6>=g_{2}g_{1}g_{3}g_{2}\vert 1>$, $\vert 7>=e_{1}g_{2}\vert 1>$,
$\vert 8>=e_{1}g_{3}g_{2}\vert 1>$,
$\vert 9>=g_{1}e_{2}\vert 1>$,
$\vert 10>=g_{2}g_{3}e_{1}g_{2}\vert 1>$, 
$\vert 11>=g_{1}g_{3}e_{2}\vert 1>$,
$\vert 12>=g_{2}g_{1}g_{3}e_{2}\vert 1>$,
$\vert 13>=e_{2}\vert 1>$,
$\vert 14>=e_{2}g_{1}g_{3}g_{2}\vert 1>$,
$\vert 15>=g_{3}e_{2}\vert 5>$, 
$\vert 16>=g_{2}g_{1}e_{3}g_{2}\vert 1>$, 
$\vert 17>= e_{3}g_{2}\vert 1>$,
$\vert 18>=e_{3}g_{1}g_{2}\vert 1>$, 
$\vert 19>=e_{3}e_{1}g_{2}\vert 1>$, 
$\vert 20>=g_{2}e_{1}e_{3}g_{2}\vert 1>$,
$\vert 21>=e_{2}g_{1}g_{3}e_{2}\vert 1>$.
\vskip 1cm
\noindent (8) $D_{2}(n)\times D_{2}(n)\uparrow D_{4}(n)$~ for~ $[2]\times [1^{2}]
= [31]+[211]+[2]+[1^{2}]+[0]$.
\vskip .3cm
\begin{tabbing}
333\=11111111\=2222222222222222222222222222222222222222222222222222222222222222222222222222222222\=\kill\\
\>{$\left\vert [31]_{1}\right>=$}
\>{$h_{1}\left(\vert 1>+\vert 2>+\vert 3>-{2\over{n+2}}(\vert 9>+\vert 13>+\vert 7>)\right)$}\\
\>{}\\
\>{$\left\vert [31]_{2}\right>=$}
\>{$-\sqrt{1\over{2}}h_{1}\left(\vert 1>-{1\over{2}}\vert 2>-{1\over{2}}\vert 3>
-{3\over{2}}(\vert 4>+\vert 5>)+{1\over{n+2}}(3\vert 8>\right.$}\\
\>{}\>{$\left.+3\vert 15>+3\vert 11>+\vert 13>+\vert 9>+\vert 7>)\right)$}\\
\>{}\\
\>{$\left\vert [31]_{3}\right>=$}
\>{$-\sqrt{3\over{2}}h_{1}\left({1\over{2}}(\vert 2>-\vert 3>+\vert 4>-\vert 5>-\vert 6>)
-{1\over{n+2}}(\vert 17>-\vert 18>\right.$}\\
\>{}\>{$\left.-\vert 13>+\vert 9>-\vert 15>+\vert 11>)\right)$}\\
\>{}\\
\>{$\left\vert [211]_{1}\right>=$}
\>{~~${\sqrt{3}\over{2\sqrt{2}(n-2)}}h_{2}\left(2\vert 7>-2\vert 8>-\sqrt{3}(
\vert 13>+\vert 9>-\vert 15>-\vert 11>)-\vert 14>\right.$}\\
\>{}\>{$\left.+\vert 10>+\vert 16>-\vert 12>+(n-2)(2\vert 1>-\vert 2>-\vert 3>
+\vert 4>+\vert 5>)\right)$}\\
\>{}\\
\>{$\left\vert [211]_{2}\right>=$}
\>{~~${\sqrt{2}\over{4}}h_{2}\left(3(\vert 2>-\vert 3>)
-\vert 4>+\vert 5>+2\vert 6>+{1\over{n-2}}(\vert 15>-\vert 11>\right.$}\\
\>{}\>{$\left.+\vert 12>+\vert 16>)-3(\vert 13>-\vert 9>+\vert 14>+\vert 10>)-2(\vert 17>-\vert
18>)\right)$,}\\
\>{}\\
\>{$\left\vert [211]_{3}\right>=$}
\>{~~$h_{2}\left(\vert 4>-\vert 5>+\vert 6>-{1\over{n-2}}(\vert 17>-\vert 18>+\vert 15>-\vert 11>+\vert 12>+\vert 16>)
\right)$,}\\
333\=1111111111\=2222222222222222222222222222222222222222222222222222222222222222222222222222222222\=\kill\\
\>{$\left\vert [2][1][0]\right>=$}
\>{$h_{3}\left(\vert 7>+\vert 8>-{2\over{n}}\vert 19>\right)$,}\\
\>{}\\
\>{$\left\vert [2][1][2]\right>=$}
\>{$\sqrt{2\over{(n+2)(n-1)}}h_{3}\left(\vert 7>+\vert 8>-\vert 21>+\vert 20>
-{n\over{2}}(\vert 13>+\vert 9>-\vert 14>+\vert 10>-{2\over{n}}\vert 19>)\right)$,}\\
\>{}\\
\>{$\left\vert [2][1][1^{2}]\right>=$}
\>{$\sqrt{n\over{2(n-1)}}h_{3}\left(\vert 9>-\vert 13>+\vert 14>+\vert 10>
-{2\over{n}}(\vert 21>+\vert 20>)\right)$,}\\
\>{}\\
\>{$\left\vert [2][21]_{1}\right>=$}
\>{$\sqrt{2\over{3(n-2)(n-1)}}h_{3}\left(\vert 7>+\vert 8>+{(n-2)\over{n}}(\vert 21>-\vert 20>
+2\vert 19>)
-{n-1\over{2}}(\vert 15>+\vert 11>\right.$}\\
\>{}\>{$\left.+\vert 12>+\vert 16>)-(n-1)(\vert 17>+\vert 18>)-{1\over{2}}(\vert 13>
+\vert 9>+\vert 14>+\vert 10>)\right),$}\\
\>{}\\
\>{$\left\vert [2][21]_{2}\right>=$}
\>{$\sqrt{n-1\over{2(n-2)}}h_{3}\left(\vert 15>-\vert 11>-\vert 12>-\vert 16>+{2(n-2)\over{n(n-1)}}(\vert 21>+
\vert 20>)\right.$}\\
\>{}\>{$\left.-{1\over{n-1}}(\vert 13>-\vert 9>-\vert 14>-\vert 10>)\right)$,}\\
333\=111111111\=2222222222222222222222222222222222222222222222222222222222222222222222222222222222\=\kill\\
\>{$\left\vert [2][3]\right>=$}
\>{$\sqrt{n+2\over{3(n+4)}}h_{3}\left(\vert 17>+\vert 18>-\vert 15>-\vert 11>-\vert 12>+\vert 16>+
\right.$}\\
\>{}\>{$\left.{2(n+4)\over{n(n+2)}}(\vert 21>-\vert 19>-\vert 20>)+{2\over{n+2}}(\vert 13>
+\vert 9>-\vert 14>+\vert 7>+\vert 10>+\vert 8>)\right),$}\\
\>{}\\
333\=111111111111\=2222222222222222222222222222222222222222222222222222222222222222222222222222222222\=\kill\\
\>{$\left\vert [1^{2}][1][0]\right>=$}
\>{$h_{4}\left(\vert 7>-\vert 8>\right)$,}\\
\>{}\\
\>{$\left\vert [1^{2}][1][2]\right>=$}
\>{$\sqrt{2\over{(n+2)(n-1)}}h_{4}\left(\vert 7>-\vert 8>
-{n\over{2}}(\vert 13>+\vert 9>)+\vert 14>-\vert 10>\right)$,}\\
\>{}\\
\>{$\left\vert [1^{2}][1][1^{2}]\right>=$}
\>{$\sqrt{n\over{2(n-1)}}h_{4}\left(\vert 9>-\vert 14>
-\vert 10>-\vert 13>\right)$,}\\
\>{}\\
\>{$\left\vert [1^{2}][21]_{1}\right>=$}
\>{$\sqrt{n(n-1)\over{2(n^{2}-4)}}h_{4}\left(\vert 15>+\vert 11>-\vert 12>+
\vert 16>+
{2\over{n-1}}(\vert 7>-\vert 8>)\right.$}\\
\>{}\>{$\left.-{1\over{n-1}}(\vert 13>+\vert 9>+\vert 14>
-\vert 10>)\right),$}\\
\>{}\\
\>{$\left\vert [1^{2}][21]_{2}\right>=$}
\>{$\sqrt{n(n-1)\over{6(n^{2}-4)}}h_{4}\left(\vert 15>-\vert 11>+\vert 12>+\vert 16>-2\vert 17>+2\vert 18>
\right.$}\\
\>{}\>{$\left.-{3\over{n-1}}(\vert 13>
-\vert 9>+\vert 14>+\vert 10>)\right),$}\\
\>{}\\
\>{$\left\vert [1^{2}][1^{3}]\right>=$}
\>{$-{1\over{3}}\sqrt{2n\over{3(n-1)}}h_{4}\left(\vert 17>-\vert 18>+\vert 15>
-\vert 11>+\vert 12>+\vert 16>\right)$,}\\
333\=111111111111\=2222222222222222222222222222222222222222222222222222222222222222222222222222222222\=\kill\\
\>{$\left\vert [0][1][0]\right>=$}\>{$h_{5}\vert 9>$,}\\
\>{}\\
\>{$\left\vert [0][1][2]\right>=$}\>{$\sqrt{2\over{(n+2)(n-1)}}h_{5}
\left({n\sqrt{n}\over{2}}(\vert 21>-\vert 20>)+\vert 19>\right)$,}\\
\>{}\\
\>{$\left\vert [0][1][1^{2}]\right>=$}\>{$\sqrt{n\over{2(n-1)}}h_{5}
\left(\vert 20>+\vert 21>\right)$,}\\
\end{tabbing}
\noindent where  $\vert 1>=\left\vert 12,
\begin{array}{l}
3\\
4
\end{array}\right>$, $\vert 2>=g_{2}\vert 1>$,
$\vert 3>=g_{1}g_{2}\vert 1>$, $\vert 4>=g_{3}g_{2}\vert 1>$,
$\vert 5>=g_{1}g_{3}g_{2}\vert 1>$,
$\vert 6>=g_{2}g_{1}g_{3}g_{2}\vert 1>$, $\vert 7>=e_{1}g_{2}\vert 1>$,
$\vert 8>=e_{1}g_{3}g_{2}\vert 1>$,
$\vert 9>=g_{1}e_{2}\vert 1>$,
$\vert 10>=g_{2}g_{3}e_{1}g_{2}\vert 1>$, 
$\vert 11>=g_{1}g_{3}e_{2}\vert 1>$,
$\vert 12>=g_{2}g_{1}g_{3}e_{2}\vert 1>$,
$\vert 13>=e_{2}\vert 1>$,
$\vert 14>=e_{2}g_{1}g_{3}g_{2}\vert 1>$,
$\vert 15>=g_{3}e_{2}\vert 5>$, 
$\vert 16>=g_{2}g_{1}e_{3}g_{2}\vert 1>$, 
$\vert 17>= e_{3}g_{2}\vert 1>$,
$\vert 18>=e_{3}g_{1}g_{2}\vert 1>$, 
$\vert 19>=e_{3}e_{1}g_{2}\vert 1>$, 
$\vert 20>=g_{2}e_{1}e_{3}g_{2}\vert 1>$,
$\vert 21>=e_{2}g_{1}g_{3}e_{2}\vert 1>$.
\vskip 1cm
\noindent (9) $D_{2}(n)\times D_{2}(n)\uparrow D_{4}(n)$~ for~ $[1^{2}]\times [1^{2}]
=[1^{4}]+[211]+[22]+[2]+[1^{2}]+[0]$.
\vskip .3cm
\begin{tabbing}
333\=11111111\=2222222222222222222222222222222222222222222222222222222222222222222222222222222222\=\kill\\
\>{$\left\vert [1^{3}]\right>=$}
\>{$k_{1}\left(\vert 1>-\vert 2>+\vert 3>+\vert 4>+\vert 6>-\vert 5>\right)$}\\
\>{}\\
\>{$\left\vert [211]_{1}\right>=$}
\>{$-\sqrt{3\over{2}}k_{2}\left(\vert 2>+\vert 3>-\vert 4>-\vert 5>
-{1\over{n-2}}(\vert 9>-\vert 15>-\vert 11>\right.$}\\
\>{}\>{$\left.+\vert 10>+\vert 12>+\vert 13>+\vert 14>-\vert 16>+2\vert 7>-2\vert 8>)\right)$}\\
\>{}\\
\>{$\left\vert [211]_{2}\right>=$}
\>{${-\sqrt{2}\over{2}}k_{2}\left(2\vert 1>-2\vert 6>+\vert 2>-\vert 3>+\vert 4>-\vert 5>
-{3\over{n-2}}(\vert 13>-\vert 9>\right.$}\\
\>{}\>{$\left.-\vert 14>-\vert 10>)
+{1\over{n-2}}(\vert 15>-\vert 11>+\vert 12>+\vert 16>-2\vert 17>+2\vert 18>)\right)$,}\\
\>{}\\
\>{$\left\vert [211]_{3}\right>=$}
\>{$-k_{2}\left(\vert 1>-\vert 2>+\vert 3>-\vert 4>+\vert 5>-\vert
6>+{2\over{n-2}}(\vert 17>-\vert 18>\right.$}\\
\>{}\>{$\left.+\vert 15> -\vert 11>+\vert 12>+\vert
16>)\right)$,}\\
\>{}\\
\>{$\left\vert [22]_{1}\right>=$}
\>{${\sqrt{3}\over{n-2}}k_{3}\left((n-2)(\vert 2>+\vert 3>+\vert 4>+\vert 5>)
-\vert 13>-\vert 9>-\vert 15>-\vert 11>\right.$}\\
\>{}\>{$\left.-\vert 14>+\vert 10>-\vert 12>+\vert 16>-2(\vert 17>+\vert
18>\right.$}\\
\>{}\>{$\left.+\vert 7>+\vert 8>)+{2\over{n-1}}(2\vert 19>+\vert 21>-\vert
20>)\right)$,}\\
\>{}\\
\>{$\left\vert [22]_{2}\right>=$}
\>{$k_{3}\left(2\vert 1>+\vert 2>-\vert 3>-\vert 4>
+\vert 5>+2\vert 6>-
{3\over{n-2}}(\vert 13>-\vert 9>-\vert 15>\right.$}\\
\>{}\>{$\left.+\vert 11>+\vert 14>+\vert 10>+\vert 16>+\vert 12>)
+{6\over{(n-2)(n-1)}}(\vert 20>+\vert 21>)\right )$}\\
333\=1111111111\=2222222222222222222222222222222222222222222222222222222222222222222222222222222222\=\kill\\
\>{$\left\vert [2][1][0]\right>=$}
\>{$k_{4}\left(\vert 7>+\vert 8>-{2\over{n}}\vert 19>\right)$,}\\
\>{}\\
\>{$\left\vert [2][1][2]\right>=$}
\>{$\sqrt{2\over{(n+2)(n-1)}}k_{4}\left(\vert 7>+\vert 8>-\vert 21>+\vert 20>
+{n\over{2}}(\vert 13>+\vert 9>-\vert 14>-\vert 10>-{2\over{n}}\vert 19>)\right)$,}\\
\>{}\\
\>{$\left\vert [2][1][1^{2}]\right>=$}
\>{$\sqrt{n\over{2(n-1)}}k_{4}\left(-\vert 9>+\vert 13>+\vert 14>+\vert 10>
-{2\over{n}}(\vert 21>+\vert 20>)\right)$,}\\
\>{}\\
\>{$\left\vert [2][21]_{1}\right>=$}
\>{$\sqrt{1\over{6(n-2)(n-1)}}k_{4}\left(\vert 13>+\vert 9>+\vert 14>-\vert 10>
+(n-1)(2\vert 17>+2\vert 18>\right.$}\\
\>{}\>{$\left.+\vert 15>+\vert 11>+\vert 12>-\vert 16>)+
2\vert 7>+2\vert 8>-2(2\vert 19>+\vert 21>-\vert 20>)\right),$}\\
\>{}\\
\>{$\left\vert [2][21]_{2}\right>=$}
\>{$\sqrt{1\over{2(n-2)(n-1)}}k_{4}\left(\vert 13>-\vert 9>+\vert 14>+\vert
10>-(n-1)(\vert 15>-\vert 11>-\vert 12>\right.$}\\
\>{}\>{$\left.-\vert 16>)-2\vert 21>-2\vert
20>\right)$}\\
\>{}\\
333\=111111111\=2222222222222222222222222222222222222222222222222222222222222222222222222222222222\=\kill\\
\>{$\left\vert [2][3]\right>=$}
\>{$-\sqrt{2\over{3(n+2)(n+4)}}k_{4}\left(\vert 13>+\vert 9>-\vert 7>+\vert 14>-\vert
10>-\vert 8>-\vert 19>-\vert 20>\right.$}\\
\>{}\>{$\left.+\vert 21>+{n+2\over{2}}(\vert 17>+\vert 18>-\vert 15>-\vert 11>-\vert
12>+\vert 16>)\right),$}\\
\>{}\\
333\=111111111111\=2222222222222222222222222222222222222222222222222222222222222222222222222222222222\=\kill\\
\>{$\left\vert [1^{2}][1][0]\right>=$}
\>{$k_{5}\left(\vert 7>-\vert 8>\right)$,}\\
\>{}\\
\>{$\left\vert [1^{2}][1][2]\right>=$}
\>{${n\over{\sqrt{2(n+2)(n-1)}}}k_{5}\left(
\vert 13>+\vert 9>-\vert 14>+\vert 10>+{2\over{n}}(\vert 7>-\vert 8>)\right)$,}\\
\>{}\\
\>{$\left\vert [1^{2}][1][1^{2}]\right>=$}
\>{$\sqrt{n\over{2(n-1)}}k_{5}\left(\vert 13>-\vert 14>
-\vert 10>-\vert 9>\right)$,}\\
\>{}\\
\>{$\left\vert [1^{2}][21]_{1}\right>=$}
\>{$\sqrt{n\over{2(n-1)(n^{2}-4)}}k_{5}\left(\vert 13>+\vert 9>-\vert 14>+\vert 10>-(n-1)(\vert
15>\right.$}\\
\>{$\left.+\vert 11>-\vert 12>+\vert 16>)+2(\vert 7>-\vert 8>)\right)$}\\
\>{}\\
\>{$\left\vert [1^{2}][21]_{2}\right>=$}
\>{$\sqrt{3n\over{2(n-1)(n^{2}-4)}}k_{5}\left(\vert 13>-\vert 9>-\vert 14>-\vert
10>\right.$}\\
\>{}\>{$\left.-{n-1\over{3}}(\vert 15>-\vert 11>+\vert 12>+\vert 16>-2\vert 17>+2\vert 18>)
\right)$}\\
\>{}\\
\>{$\left\vert [1^{2}][1^{3}]\right>=$}
\>{$\sqrt{2n(n-1)\over{9(n-2)}}k_{5}\left(\vert 17>-\vert 18>+\vert 15>
-\vert 11>+\vert 12>+\vert 16>\right)$,}\\
333\=111111111111\=2222222222222222222222222222222222222222222222222222222222222222222222222222222222\=\kill\\
\>{$\left\vert [0][1][0]\right>=$}\>{$k_{6}\vert 9>$,}\\
\>{}\\
\>{$\left\vert [0][1][2]\right>=$}\>{${n\over{\sqrt{2(n+2)(n-1)}}}k_{6}
\left( \vert 21>-\vert 20>-{2\over{n}}\vert 19>\right)$,}\\
\>{}\\
\>{$\left\vert [0][1][1^{2}]\right>=$}\>{${n\over{2(n-1)}}k_{6}
\left( \vert 20>+\vert 21>\right)$,}\\
\end{tabbing}
\vskip .5cm
\noindent where  $\vert 1>=\left\vert 
\begin{array}{l}
1\\
2
\end{array},
\begin{array}{l}
3\\
4
\end{array}\right>$, $\vert 2>=g_{2}\vert 1>$, 
$\vert 3>=g_{1}g_{2}\vert 1>$, $\vert 4>=g_{3}g_{2}\vert 1>$, $\vert 5>=g_{1}g_{3}g_{2}\vert 1>$,
$\vert 6>=g_{2}g_{1}g_{3}g_{2}\vert 1>$, $\vert 7>=e_{1}g_{2}\vert 1>$,
$\vert 8>=e_{1}g_{3}g_{2}\vert 1>$,
$\vert 9>=g_{1}e_{2}\vert 1>$,
$\vert 10>=g_{2}g_{3}e_{1}g_{2}\vert 1>$, 
$\vert 11>=g_{1}g_{3}e_{2}\vert 1>$,
$\vert 12>=g_{2}g_{1}g_{3}e_{2}\vert 1>$,
$\vert 13>=e_{2}\vert 1>$,
$\vert 14>=e_{2}g_{1}g_{3}g_{2}\vert 1>$,
$\vert 15>=g_{3}e_{2}\vert 5>$, 
$\vert 16>=g_{2}g_{1}e_{3}g_{2}\vert 1>$, 
$\vert 17>= e_{3}g_{2}\vert 1>$,
$\vert 18>=e_{3}g_{1}g_{2}\vert 1>$, 
$\vert 19>=e_{3}e_{1}g_{2}\vert 1>$, 
$\vert 20>=g_{2}e_{1}e_{3}g_{2}\vert 1>$,
$\vert 21>=e_{2}g_{1}g_{3}e_{2}\vert 1>$.
\vskip .5cm
   In the above expressions, the IDCs are determined up to a  absolute normalization constant,
which can be calculated for different cases by using the corresponding norm matrix
as has been done in the previous example.  In the next section, we will outline an
assimilation method for evaluating CGCs of $O(n)$ in the Gel'fand basis from these
IDCs. 
\vskip .5cm
\begin{centering}
{\large 4. Evaluating CGCs of $SO(n)$ in its Gel'fand basis}\\
\end{centering}
\vskip .3cm
   Irreducible representations of $SO(m)$, where $m=2,3,\cdots n$, can be 
labeled by partitions $[\lambda_{1m}\lambda_{2m}\cdots\lambda_{hm}]$
with $h=m/2$ for $m$ even, and $h=(m-1)/2$ for $m$ odd, which satisfy
the condition

$$\lambda_{1m}\geq \lambda_{2m}\geq\cdots\geq\lambda_{hm}\geq0~~{\rm for~ m~ odd},$$

$$\lambda_{1m}\geq \lambda_{2m}\geq\cdots\geq\vert\lambda_{hm}\vert\geq0~~{\rm for~ m~ even},\eqno(29)$$

\noindent where $\lambda_{im}$ ($i=1,2,\cdots,h$) are all integers
because we only discuss tensor representations of $SO(m)$. 
The partitions of two groups, for example, $SO(2p+1)$ and $SO(2p)$,
in the canonical chain $SO(n)\supset\cdots\supset SO(2p+1)\supset
SO(2p)\supset\cdots\supset SO(2)$ are related by the betweeness
conditions

$$\lambda_{1~2p+1}\geq\lambda_{1~2p}\geq\lambda_{2~2p+1}\geq\lambda_{2~2p}\geq\cdots\geq\lambda_{p~2p+1}
\geq\vert\lambda_{p~2p}\vert.\eqno(30)$$

   Similar to $U(n)$ case, we can define  Weyl tableau for $SO(n)$ in the Gel'fand basis. 
\vskip .3cm
$$W^{[\lambda ]}=
\begin{array}{l}
{\begin{tabular}{|l|l|l|l|l|l|}
\hline
$f_{12}~(\pm a_{2})$'s{~} &$f_{13}~a_{3}$'s{~} &$f_{14}~a_{4}$'s &{$f_{15}~a_{5}$'s}&$f_{16}~a_{6}$'s &$\cdots$\\
\hline
\end{tabular}}\\
{\begin{tabular}{|l|l|l|l|}
$~~~f_{24}~(\pm a_{4})~~$'s &$f_{25}~a_{5}$'s &{$f_{26}~a_{6}$'s} &$\cdots$\\
\hline
\end{tabular}}\\
{\begin{tabular}{|l|l|}
$~~~~f_{36}~(\pm a_{6})$'s~~~ &$\cdots\cdots$\\
\hline
\end{tabular}}\\
{\begin{tabular}{|l|}
$\cdots\cdots$\\
\hline
\end{tabular}}
\end{array}\eqno(31)
$$
 where the signs in the front of $a_{2k}$ ($k=1,2,\cdots$)
should always be the same. They can be taken all positive or all negative. 
The correspondence between the Weyl tableau and the Gel'fand basis is  realized in the following way:

$$\pm f_{12}=\lambda_{12},~~f_{12}+f_{13}=\lambda_{13},~~f_{12}+f_{13}+f_{14}=\lambda_{14},\cdots,$$

$$\pm f_{24}=\lambda_{24},~~f_{24}+f_{25}=\lambda_{25},~~f_{24}+f_{25}+f_{26}=\lambda_{26},\cdots,$$

$$\pm f_{36}=\lambda_{36},~~f_{36}+f_{46}=\lambda_{46},~~\cdots.\eqno(32)$$

\vskip .3cm
\noindent For example, basis vectors of $SO(5)\supset SO(4)\supset SO(3)\supset SO(2)$ can be
denoted either by Gel'fand  symbol or by Weyl tableau as

$$\left(
\matrix{
[\lambda_{15}\lambda_{25}]\cr
[\lambda_{14}\lambda_{24}]\cr
\lambda_{13}\cr
\lambda_{12}\cr
}\right)=\begin{array}{l}
{\begin{tabular}{|l|l|l|l|l|l|}
\hline
$f_{12}~(\pm a_{2})$'s{~} &$f_{13}~a_{3}$'s{~} &$f_{14}~a_{4}$'s &{$f_{15}~a_{5}$'s}\\
\hline
\end{tabular}}\\
{\begin{tabular}{|l|l|l|l|}
$~~~f_{24}~(\pm a_{4})~~$'s &$f_{25}~a_{5}$'s \\
\hline
\end{tabular}}
\end{array},\eqno(33)$$

\noindent where

$$f_{12}=\vert\lambda_{12}\vert,~~f_{13}=\lambda_{13}-\vert\lambda_{12}\vert,~~f_{14}=\lambda_{14}-\lambda_{13},~~
f_{15}=\lambda_{15}-\lambda_{14},$$

$$f_{24}=\vert\lambda_{24}\vert,~~f_{25}=\lambda_{25}-\vert\lambda_{24}\vert,\eqno(34)$$

\noindent and  signs in the front of $a_{2}$ or $a_{4}$ should be taken as positive (negative)
if $\lambda_{12}\geq 0$ ($<
 0$) or $\lambda_{24}\geq 0$ ($< 0$).

   An assimilation method for obtaining CGCs of $SO(n)$ from IDCs of Brauer algebra is the following:
Firstly, the one-box representation of $D_{1}(n)$ is just rank-1 tensor of $SO(n)$

$$\begin{tabular}{|l|}
\hline
i\\
\hline
\end{tabular}\rightarrow T^{i}_{j_{i}},\eqno(35)$$

 \noindent where the index $i$ is used to indicate that the tensor operator is in the $i$-th space,
while $j_{i}$ is the tensor component, and can be taken as $n$ different values. Then, an irrep
$[\lambda ]$ of $SO(n)$ can be constructed by rank-1 tensors through cabling

$$T^{1}_{i_{1}}T^{2}_{i_{2}}\cdots T^{f}_{i_{f}}\Longrightarrow T^{[\lambda]_{f}}_{i_{1}i_{2}\cdots i_{f}}.\eqno(36)$$

\noindent Next, the symmetry properties of $f$ space indices $\{ 1,2,\cdots,f\}$ and those tensor components
$\{ i_{1},i_{2},\cdots,i_{f}\}$ are the same, i.e. interchange of $i$ and $k$ is the same as interchange
of $j_{i}$ and $j_{k}$. Hence, there is a natural assimilation

$$i\longrightarrow j_{i}.\eqno(37)$$

\noindent After interchange $i$ with $j_{i}$ in   basis vectors  of $S_{f_{1}}\times S_{f_{2}}\uparrow D_{f}(n)$,
 the resulting basis vectors become the corresponding $SO(n)$ orthogonal basis vectors in Weyl tableau forms.  This fact
just reflects the Brauer-Schur-Weyl duality relation between $D_{f}(n)$ and $O(n)$. For example, the
basis vector of $D_{1}(n)\times
D_{1}(n)\uparrow D_{2}(n)$ induced representation for the coupling $[1]\times [1]\uparrow [1^{2}]$
can be expressed as

$$\left\vert
\begin{array}{l}
{\begin{tabular}{|l|}
\hline
$1$ \\
\hline
\end{tabular}}\\
{\begin{tabular}{|l|l|l|l|}
$2$ \\
\hline
\end{tabular}}
\end{array}\right>
=\sqrt{1\over{2}}(\left\vert
\begin{array}{l}
{\begin{tabular}{|l|}
\hline
$1$\\
\hline
\end{tabular}}
\end{array}, \begin{array}{l}
\begin{tabular}{|l|}
\hline
$2$\\
\hline
\end{tabular}
\end{array}\right>-\left\vert
\begin{array}{l}
\begin{tabular}{|l|}
\hline
$2$\\
\hline
\end{tabular}
\end{array}, \begin{array}{l}
\begin{tabular}{|l|}
\hline
$1$\\
\hline
\end{tabular}
\end{array}\right>).\eqno(38)$$

\noindent After the assimilation, one get the corresponding orthogonal basis  vector of $SO(n)\times SO(n)\rightarrow SO(n)$
coupling with

$$\left\vert
\begin{array}{l}
{\begin{tabular}{|l|}
\hline
$i_{1}$ \\
\hline
\end{tabular}}\\
{\begin{tabular}{|l|l|l|l|}
$i_{2}$ \\
\hline
\end{tabular}}
\end{array}\right>=\sqrt{1\over{2}}(\left\vert
\begin{array}{l}
\begin{tabular}{|l|}
\hline
$i_{1}$\\
\hline
\end{tabular}
\end{array}, \begin{array}{l}
\begin{tabular}{|l|}
\hline
$i_{2}$\\
\hline
\end{tabular}
\end{array}\right>-\left\vert
\begin{array}{l}
\begin{tabular}{|l|}
\hline
$i_{2}$\\
\hline
\end{tabular}
\end{array}, \begin{array}{l}
\begin{tabular}{|l|}
\hline
$i_{1}$\\
\hline
\end{tabular}
\end{array}\right>).\eqno(39)$$

\noindent The left hand sides of (38) and (39) are the same, namely

$$\sqrt{1\over{2}}(\left\vert
\begin{array}{l}
\begin{tabular}{|l|}
\hline
$1$\\
\hline
\end{tabular}
\end{array}, \begin{array}{l}
\begin{tabular}{|l|}
\hline
$2$\\
\hline
\end{tabular}
\end{array}\right>-\left\vert
\begin{array}{l}
\begin{tabular}{|l|}
\hline
$2$\\
\hline
\end{tabular}
\end{array}, \begin{array}{l}
\begin{tabular}{|l|}
\hline
$1$\\
\hline
\end{tabular}
\end{array}\right>)=\sqrt{1\over{2}}(\left\vert
\begin{array}{l}
\begin{tabular}{|l|}
\hline
$i_{1}$\\
\hline
\end{tabular}
\end{array},
 \begin{array}{l}
\begin{tabular}{|l|}
\hline
$i_{2}$\\
\hline
\end{tabular}
\end{array}\right>-\left\vert
\begin{array}{l}
\begin{tabular}{|l|}
\hline
$i_{2}$\\
\hline
\end{tabular}
\end{array}, \begin{array}{l}
{\begin{tabular}{|l|}
\hline
$i_{1}$\\
\hline
\end{tabular}}
\end{array}\right>)=$$

$$\sqrt{1\over{2}}(1-g_{1})\vert 1,2>\equiv \sqrt{1\over{2}}(
T^{12}_{i_{1}i_{2}}-T^{12}_{i_{2}i_{1}}).\eqno(40)$$

\noindent The difference is only the space indices are interchanged to
the corresponding $SO(n)$ tensor components. Furthermore, such interchange
keeps Brauer algebra action $g_{i}$ or $e_{i}$ on the uncoupled
basis vectors unchanged. 
While the meaning of
(38) and (39) is different. The former gives the basis vector of induced representation
of $D_{1}(n)\times D_{1}(n)\uparrow D_{2}(n)$, the latter gives
basis vector of $SO(n)\times SO(n)\rightarrow SO(n)$  in the canonical basis. This
assimilation can thus be obtained just because of the Brauer-Schur-Weyl duality
between $D_{f}(n)$ and $O(n)$. Furthermore, the phase convention 
of $SO(n)$ CGCs
have already been determined by that of IDCs of Brauer algebra. Therefore, it is
not necessary to consider the phase convention for $SO(n)$ CGCs  again.
\vskip .3cm
   However, according to Lemma 2 of Ref. [11], for the group $O(n)$, where $n=2l$ or
$2l+1$, the irrep $[\lambda_{1n},\lambda_{2n},\cdots,\lambda_{pn},\dot{0}]$ 
is non-standard
if $p>l$. In these cases, modification rules will be needed. In such circumstances,
some irregular representations  will involved, which can not be obtained by using
the assimilation method. For example, coupled basis vectors of
$[1]\times [1]\rightarrow [11]$ or $[1~-1]$ for $SO(4)$ can not be expressed  by uncoupled
basis vectors as given by (39) because the subirreps of SO(4) involve the coupling
$[1]\times [1]\rightarrow [11]$ for SO(3). The irrep $[11]=[1]$ for SO(3) obviously 
needs modification rules for $O(n)$. i.e., (39) is only valid for $SO(n)$ for $n\geq 5$.  
Therefore, we only consider CGCs with no modification rule needed couplings.
 Using this assimilation method, one can evaluate
CGCs of $SO(n)$ in its canonical basis  with no modification needed couplings 
from IDCs of $S_{f_{1}}\times S_{f_{2}}\uparrow
D_{f}(n)$.  In the following, we will give an example  to show how this method works.
\vskip .3cm
\noindent {\bf Example 2}. Find SO(n) CGCs for $[1]\times [1]=[2]+[1^{2}]^{*}+[0]$, where
$*$ indicates this irrep is only valid for $n\geq 5$.
\vskip .3cm
\noindent Step 1. Write the corresponding basis vectors of $S_{1}\times S_{1}\uparrow
D_{2}(n)$. Using the results given in the previous section, we have

$$\left\vert
\begin{array}{l}
{\begin{tabular}{|l|l|}
\hline
$1$ &$2$ \\
\hline
\end{tabular}}\\
\end{array}\right>
=a_{1}(1+g_{1}-{2\over{n}}e_{1})\left\vert
\begin{array}{l}
{\begin{tabular}{|l|}
\hline
$1$\\
\hline
\end{tabular}}
\end{array}, \begin{array}{l}
\begin{tabular}{|l|}
\hline
$2$\\
\hline
\end{tabular}
\end{array}\right>,\eqno(41a)$$

$$\left\vert
\begin{array}{l}
{\begin{tabular}{|l|}
\hline
$1$ \\
\hline
\end{tabular}}\\
{\begin{tabular}{|l|l|l|l|}
$2$ \\
\hline
\end{tabular}}
\end{array}\right>=\sqrt{1\over{2}}(1-g_{1})\left\vert
\begin{array}{l}
\begin{tabular}{|l|}
\hline
$1$\\
\hline
\end{tabular}
\end{array}, \begin{array}{l}
\begin{tabular}{|l|}
\hline
$2$\\
\hline
\end{tabular}
\end{array}\right>,\eqno(41b)$$

$$\vert [0]>=\sqrt{1\over{n}}e_{1}\left\vert
\begin{array}{l}
\begin{tabular}{|l|}
\hline
$1$\\
\hline
\end{tabular}
\end{array}, \begin{array}{l}
\begin{tabular}{|l|}
\hline
$2$\\
\hline
\end{tabular}
\end{array}\right>.\eqno(41c)$$
\vskip .3cm
\noindent Step 2. Make assimilations. We need  consider three different
ways of assimilation in this case.
\vskip .3cm
\noindent (a)  $i_{1}=\tau\alpha_{k}$, $i_{2}=\alpha_{m}$ ($k<m$),
where $\tau$ can be taken as $-$ for $k=2$, and $\tau=+$ for other cases.  Because
the tensor indices are different, contraction of $i_{1}$ and $i_{2}$ is zero.
Hence, we get

$$\left\vert
\begin{array}{l}
{\begin{tabular}{|l|l|}
\hline
$\tau\alpha_{k}$ &$\alpha_{m}$ \\
\hline
\end{tabular}}\\
\end{array}\right>
=\sqrt{1\over{2}}(\left\vert
\begin{array}{l}
{\begin{tabular}{|l|}
\hline
$\tau\alpha_{k}$\\
\hline
\end{tabular}}
\end{array}, \begin{array}{l}
\begin{tabular}{|l|}
\hline
$\alpha_{m}$\\
\hline
\end{tabular}
\end{array}\right>+\left\vert
\begin{array}{l}
{\begin{tabular}{|l|}
\hline
$\alpha_{m}$\\
\hline
\end{tabular}}
\end{array}, \begin{array}{l}
\begin{tabular}{|l|}
\hline
$\tau\alpha_{k}$\\
\hline
\end{tabular}
\end{array}\right>),\eqno(42a)$$

$$\left\vert
\begin{array}{l}
{\begin{tabular}{|l|}
\hline
$\tau\alpha_{k}$ \\
\hline
\end{tabular}}\\
{\begin{tabular}{|l|l|l|l|}
$\delta\alpha_{m}$ \\
\hline
\end{tabular}}
\end{array}\right>=\sqrt{1\over{2}}(\left\vert
\begin{array}{l}
{\begin{tabular}{|l|}
\hline
$\tau\alpha_{k}$\\
\hline
\end{tabular}}
\end{array}, \begin{array}{l}
\begin{tabular}{|l|}
\hline
$\alpha_{m}$\\
\hline
\end{tabular}
\end{array}\right>-\left\vert
\begin{array}{l}
{\begin{tabular}{|l|}
\hline
$\alpha_{m}$\\
\hline
\end{tabular}}
\end{array}, \begin{array}{l}
\begin{tabular}{|l|}
\hline
$\tau\alpha_{k}$\\
\hline
\end{tabular}
\end{array}\right>),\eqno(42b)$$

\noindent where the normalization factor $a_{1}=\sqrt{1\over{2}}$.  It should be noted
that the SO(2m) tensor indices $\delta\alpha_{2m}$ for $n$ even can be taken $-\alpha_{2m}$  
only in the $m$-th row in the Weyl tableau. $-\alpha_{2m}$ in the $p$-th row with $p<m$
is forbidden according to the definition of $SO(n)$ Weyl tableau. Hence, $\delta\alpha_{m}$
should be replaced by $\alpha_{m}$ in the $p$-th row with $p<m$. Thus, $\delta$ can be taken as
$-$ only for $m=4$, and the only possible minus sign of $\tau$ allowed in the first row
is $k=2$ case. For example, from (42a), for $m=3$ and $k=2$,  one gets the following
CGCs of $SO(3)$

$$\left<
\matrix{
[1] &[1]\cr
\pm 1 &0\cr}\right\vert\left.
\matrix{
[2] \cr
\pm 1\cr}\right>=\sqrt{1\over{2}},~~\left<
\matrix{
[1] &[1]\cr
0 &\pm 1\cr}\right\vert\left.
\matrix{
[2] \cr
\pm 1 \cr}\right>=\sqrt{1\over{2}}.\eqno(43a)$$

\noindent From (42b) for $k=3$ and $m=5$, one gets the following CGCs for $SO(5)$
 
$$\left(
\matrix{
[1] &[1]\cr
[1] &0\cr
[1] &0\cr
0 &0\cr
}\right\vert\left.
\matrix{
[11] \cr
[1]\cr
[1]\cr
0\cr}\right)=\sqrt{1\over{2}},~~\left(
\matrix{
[1] &[1]\cr
[0] &[1]\cr
[0] &[1]\cr
0  & 0\cr}\right\vert\left.
\matrix{
[11] \cr
[1]\cr
[1]\cr
0 \cr}\right)=-\sqrt{1\over{2}}.\eqno(43b)$$

\noindent For $m\leq 4$ in (42b) there will be representations involving modification
rules, which will not be considered in this paper. 

\vskip .3cm
\noindent (b) $i_{1}=i_{2}=a_{n}$. In this case, the trace contraction is nonzero.
We have

$$e_{1}\left\vert
\begin{array}{l}
{\begin{tabular}{|l|}
\hline
$\alpha_{n}$\\
\hline
\end{tabular}}
\end{array}, \begin{array}{l}
\begin{tabular}{|l|}
\hline
$\alpha_{n}$\\
\hline
\end{tabular}
\end{array}\right>=\sum_{\mu}{}^{(s)}\left\vert
\begin{array}{l}
{\begin{tabular}{|l|}
\hline
$\alpha_{\mu}$\\
\hline
\end{tabular}}
\end{array}, \begin{array}{l}
\begin{tabular}{|l|}
\hline
$\alpha_{\mu}$\\
\hline
\end{tabular}
\end{array}\right>,\eqno(44a)$$

\noindent where there are $n$ terms involved in the sum.
It should be understood that the sum on the right hand 
side of (44a) is shorthand notation, of which the exact
expression should be 

$$\sum_{\mu}{}^{(s)}\left\vert
\begin{array}{l}
{\begin{tabular}{|l|}
\hline
$\alpha_{\mu}$\\
\hline
\end{tabular}}
\end{array}, \begin{array}{l}
\begin{tabular}{|l|}
\hline
$\alpha_{\mu}$\\
\hline
\end{tabular}
\end{array}\right>=\sum_{\mu\geq 3}\left\vert
\begin{array}{l}
{\begin{tabular}{|l|}
\hline
$\alpha_{\mu}$\\
\hline
\end{tabular}}
\end{array}, \begin{array}{l}
\begin{tabular}{|l|}
\hline
$\alpha_{\mu}$\\
\hline
\end{tabular}
\end{array}\right>
+\sum_{\delta=+,-}(-)
\left\vert
\begin{array}{l}
{\begin{tabular}{|l|}
\hline
$\delta\alpha_{2}$\\
\hline
\end{tabular}}
\end{array}, \begin{array}{l}
\begin{tabular}{|l|}
\hline
$-\delta\alpha_{2}$\\
\hline
\end{tabular}
\end{array}\right>.\eqno(44b)$$

\noindent Thus, we get

$$\left\vert
\begin{array}{l}
{\begin{tabular}{|l|l|}
\hline
$\alpha_{n}$ &$\alpha_{n}$ \\
\hline
\end{tabular}}\\
\end{array}\right>=2a_{1}\left(
(1-{1\over{n}})\left\vert
\begin{array}{l}
{\begin{tabular}{|l|}
\hline
$\alpha_{n}$\\
\hline
\end{tabular}}
\end{array}, \begin{array}{l}
\begin{tabular}{|l|}
\hline
$\alpha_{n}$\\
\hline
\end{tabular}
\end{array}\right>-{1\over{n}}\sum_{\mu\neq n}{}^{(s)}\left\vert
\begin{array}{l}
{\begin{tabular}{|l|}
\hline
$\alpha_{\mu}$\\
\hline
\end{tabular}}
\end{array}, \begin{array}{l}
\begin{tabular}{|l|}
\hline
$\alpha_{\mu}$\\
\hline
\end{tabular}
\end{array}\right>\right).\eqno(45)$$

\noindent After normalization, (45) becomes

$$\left\vert
\begin{array}{l}
{\begin{tabular}{|l|l|}
\hline
$\alpha_{n}$ &$\alpha_{n}$ \\
\hline
\end{tabular}}\\
\end{array}\right>=\sqrt{n-1\over{n}}
\left\vert
\begin{array}{l}
{\begin{tabular}{|l|}
\hline
$\alpha_{n}$\\
\hline
\end{tabular}}
\end{array}, \begin{array}{l}
\begin{tabular}{|l|}
\hline
$\alpha_{n}$\\
\hline
\end{tabular}
\end{array}\right>-\sqrt{1\over{n(n-1)}}\sum_{\mu\neq n}{}^{(s)}\left\vert
\begin{array}{l}
{\begin{tabular}{|l|}
\hline
$\alpha_{\mu}$\\
\hline
\end{tabular}}
\end{array}, \begin{array}{l}
\begin{tabular}{|l|}
\hline
$\alpha_{\mu}$\\
\hline
\end{tabular}
\end{array}\right>.\eqno(46)$$

\noindent Similarly, we have

$$\vert [0]>=\sqrt{1\over{n}}
\left\vert
\begin{array}{l}
{\begin{tabular}{|l|}
\hline
$\alpha_{n}$\\
\hline
\end{tabular}}
\end{array}, \begin{array}{l}
\begin{tabular}{|l|}
\hline
$\alpha_{n}$\\
\hline
\end{tabular}
\end{array}\right>+\sqrt{1\over{n}}\sum_{\mu\neq n}{}^{(s)}\left\vert
\begin{array}{l}
{\begin{tabular}{|l|}
\hline
$\alpha_{\mu}$\\
\hline
\end{tabular}}
\end{array}, \begin{array}{l}
\begin{tabular}{|l|}
\hline
$\alpha_{\mu}$\\
\hline
\end{tabular}
\end{array}\right>.\eqno(47)$$

\vskip .3cm
\noindent (c) $i_{1}=i_{2}=a_{k}$ ($2<k<n$).  The final results in this case
 are similar to those of (46) and (47). 

$$\left\vert
\begin{array}{l}
{\begin{tabular}{|l|l|}
\hline
$\alpha_{k}$ &$\alpha_{k}$ \\
\hline
\end{tabular}}\\
\end{array}\right>=\sqrt{k-1\over{k}}
\left\vert
\begin{array}{l}
{\begin{tabular}{|l|}
\hline
$\alpha_{k}$\\
\hline
\end{tabular}}
\end{array}, \begin{array}{l}
\begin{tabular}{|l|}
\hline
$\alpha_{k}$\\
\hline
\end{tabular}
\end{array}\right>-\sqrt{1\over{k(k-1)}}\sum_{\mu\neq k}{}^{(s)}\left\vert
\begin{array}{l}
{\begin{tabular}{|l|}
\hline
$\alpha_{\mu}$\\
\hline
\end{tabular}}
\end{array}, \begin{array}{l}
\begin{tabular}{|l|}
\hline
$\alpha_{\mu}$\\
\hline
\end{tabular}
\end{array}\right>.\eqno(48)$$

$$\vert [0]>=\sqrt{1\over{k}}
\left\vert
\begin{array}{l}
{\begin{tabular}{|l|}
\hline
$\alpha_{k}$\\
\hline
\end{tabular}}
\end{array}, \begin{array}{l}
\begin{tabular}{|l|}
\hline
$\alpha_{k}$\\
\hline
\end{tabular}
\end{array}\right>+\sqrt{1\over{k}}\sum_{\mu\neq k}{}^{(s)}\left\vert
\begin{array}{l}
{\begin{tabular}{|l|}
\hline
$\alpha_{\mu}$\\
\hline
\end{tabular}}
\end{array}, \begin{array}{l}
\begin{tabular}{|l|}
\hline
$\alpha_{\mu}$\\
\hline
\end{tabular}
\end{array}\right>.\eqno(49)$$

\noindent  The corresponding CGCs of $SO(n)$ can now be read off from 
(42a,b), (46)-(48).  When $n=4$, for example, the $SO(4)\supset SO(3)\supset SO(2)$
 CGCs read off from (46) and (47) are

$$\left<
\matrix{
[1] &[1]\cr
[0] &[0]\cr
0 &0\cr}\right\vert\left.
\matrix{
[2] \cr
[0]\cr
0 \cr}\right>=\sqrt{3\over{4}},~~\left<
\matrix{
[1] &[1]\cr
[1] &[1]\cr
1 &-1\cr}\right\vert\left.
\matrix{
[2] \cr
[0]\cr
0 \cr}\right>=\left<
\matrix{
[1] &[1]\cr
[1] &[1]\cr
-1 &1\cr}\right\vert\left.
\matrix{
[2] \cr
[0]\cr
0 \cr}\right>=\sqrt{1\over{12}},$$

$$\left<
\matrix{
[1] &[1]\cr
[1] &[1]\cr
0 &0\cr}\right\vert\left.
\matrix{
[2] \cr
[0]\cr
0 \cr}\right>=-\sqrt{1\over{12}},~~\left<
\matrix{
[1] &[1]\cr
[0] &[0]\cr
0 &0\cr}\right\vert\left.
\matrix{
[0] \cr
[0]\cr
0 \cr}\right>=\left<
\matrix{
[1] &[1]\cr
[1] &[1]\cr
0 &0\cr}\right\vert\left.
\matrix{
[0] \cr
[0]\cr
0 \cr}\right>=\sqrt{1\over{4}},$$

$$\left<
\matrix{
[1] &[1]\cr
[1] &[1]\cr
1 &-1\cr}\right\vert\left.
\matrix{
[0] \cr
[0]\cr
0 \cr}\right>=\left<
\matrix{
[1] &[1]\cr
[1] &[1]\cr
-1 &1\cr}\right\vert\left.
\matrix{
[0] \cr
[0]\cr
0 \cr}\right>=-\sqrt{1\over{4}}.\eqno(50)$$

\noindent When $n=4$ and $k=3$, from (48) and (49), we have

$$\left<
\matrix{
[1] &[1]\cr
[1] &[1]\cr
0 &0\cr}\right\vert\left.
\matrix{
[2] \cr
[2]\cr
0 \cr}\right>=\sqrt{2\over{3}},~~\left<
\matrix{
[1] &[1]\cr
[1] &[1]\cr
1 &-1\cr}\right\vert\left.
\matrix{
[2] \cr
[2]\cr
0 \cr}\right>=\left<
\matrix{
[1] &[1]\cr
[1] &[1]\cr
-1 &1\cr}\right\vert\left.
\matrix{
[2] \cr
[2]\cr
0 \cr}\right>=\sqrt{1\over{6}},\eqno(51)$$

$$\left<
\matrix{
[1] &[1]\cr
0 &0\cr}\right\vert\left.
\matrix{
[0] \cr
0 \cr}\right>=\sqrt{1\over{3}},~~\left<
\matrix{
[1] &[1]\cr
1 &-1\cr}\right\vert\left.
\matrix{
[0] \cr
0 \cr}\right>=\left<
\matrix{
[1] &[1]\cr
-1 &1\cr}\right\vert\left.
\matrix{
[0] \cr
0 \cr}\right>=-\sqrt{1\over{3}}.\eqno(52)$$
\vskip .3cm
\noindent where (52) gives CGCs of $SO(3)$. 
\vskip .3cm
\noindent Using this method, we have  derived CGCs of $SO(n)$ for the resulting irrep
$[\lambda_{1},~\lambda_{2},~\lambda_{3},~\lambda_{4},\dot{0}]$ with
$\sum\limits_{i=1}^{4}\lambda_{i}\leq 4$
and no modification  rule involved couplings
from IDCs of $S_{f_{1}}\times
S_{f_{2}}\uparrow D_{f}(n)$ with $f_{1}+f_{2}=f$ given in the previous section. However, the 
expressions of the CGCs for any $n$ are too
cumbersome to be  tabulated. While the Isoscalar factors (ISFs) of $SO(n)\supset SO(n-1)$ for
any $n$, which can be obtained
according to Racah factorization lemma,$^{[5]}$ are concise and easily
to be listed in a table. For example,  one can easily read off
the following ISFs of $SO(n)\supset SO(n-1)$ with $n\geq 4$ from (46):
\vskip .3cm
$$\left<
\matrix{
[1] &[1]\cr
[0] &[0]\cr}\right\vert\left.
\matrix{
[2]\cr
[0]\cr}\right>=\sqrt{n-1\over{n}},~~
\left<
\matrix{
[1] &[1]\cr
[1] &[1]\cr}\right\vert\left.
\matrix{
[2]\cr
[0]\cr}\right>=-\sqrt{1\over{n}}.\eqno(53)$$
\vskip .3cm
\noindent The notation for the ISFs used in (53) is much simpler than
that of CGCs. Therefore, we shall only list ISFs of $SO(n)\supset SO(n-1)$ for
$n\geq 4$ in the next section.  
\vskip .5cm
\begin{centering}
{\large 5. ISFs of $SO(n)\supset SO(n-1)$}\\
\end{centering}
\vskip .4truecm
    In this section, we will list some ISFs of $SO(n)\supset SO(n-1)$ derived
by using the assimilation method outlined in Sec. 4. According to
Racah factorization lemma, $SO(n)$ CGCs in the canonical basis
$SO(n)\supset SO(n-1)\supset\cdots\supset SO(2)$ can be expressed as

$$\left(\matrix{
[\lambda_{1n}] &[\lambda_{2n}]\cr
[\lambda_{1~n-1}] &[\lambda_{2~n-1}]\cr
\cdots&\cdots\cr
 m_{12} &m_{22}\cr}\right.\left\vert
\matrix{\tau_{n}~[\lambda_{n}]\cr
~~~[\lambda_{n-1}]\cr
\cdots\cr
{~~~[m_{2}]}\cr}\right)=\sum_{\tau_{n-1}}
\left<\matrix{
[\lambda_{1n}] &[\lambda_{2n}]
\cr
[\lambda_{1~n-1}] &[\lambda_{2~n-1}]\cr}\right.\left\vert
\matrix{\tau_{n}~[\lambda_{n}]\cr
\tau_{n-1}~[\lambda_{n-1}]\cr}\right>\times$$

$$\left(\matrix{
[\lambda_{1~n-1}] &[\lambda_{2~n}]\cr
[\lambda_{1~n-2}] &[\lambda_{2~n-2}]\cr
\cdots&\cdots\cr
 m_{12} &m_{22}\cr}\right.\left\vert
\matrix{\tau_{n-1}~[\lambda_{n-1}]\cr
~~~~[\lambda_{n-2}]\cr
\cdots\cr
{~~~[m_{2}]}\cr}\right),\eqno(54)$$

\noindent where 

$$\left(\matrix{
[\lambda_{1n}] &[\lambda_{2n}]\cr
[\lambda_{1~n-1}] &[\lambda_{2~n-1}]\cr
\cdots&\cdots\cr
 m_{12} &m_{22}\cr}\right.\left\vert
\matrix{\tau_{n}~[\lambda_{n}]\cr
~~~[\lambda_{n-1}]\cr
\cdots\cr
{~~~[m_{2}]}\cr}\right)\eqno(55)$$

\noindent is $SO(n)$ CGC, 

$$\left<\matrix{
[\lambda_{1n}] &[\lambda_{2n}]
\cr
[\lambda_{1~n-1}] &[\lambda_{2~n-1}]\cr}\right.\left\vert
\matrix{\tau_{n}~[\lambda_{n}]\cr
\tau_{n-1}~[\lambda_{n-1}]\cr}\right>$$

\noindent is $SO(n)\supset SO(n-1)$ ISF, and $\tau_{n}$ is the multiplicity
label needed in the coupling $[\lambda_{1n}]\times [\lambda_{2n}]\downarrow
[\lambda_{n}]$. The ISFs satisfy the following orthogonality conditions

$$\sum_{\lambda_{1~n-1}\lambda_{2~n-1}}\left<\matrix{
[\lambda_{1n}] &[\lambda_{2n}]
\cr
[\lambda_{1~n-1}] &[\lambda_{2~n-1}]\cr}\right.\left\vert
\matrix{\tau_{n}~[\lambda_{n}]\cr
\tau_{n-1}~[\lambda_{n-1}]\cr}\right>
\left<\matrix{
[\lambda_{1n}] &[\lambda_{2n}]
\cr
[\lambda_{1~n-1}] &[\lambda_{2~n-1}]\cr}\right.\left\vert
\matrix{\tau^{\prime}_{n}~[\lambda^{\prime}_{n}]\cr
\tau_{n-1}~[\lambda_{n-1}]\cr}\right>=\delta_{\lambda^{\prime}_{n}\lambda_{n}}
\delta_{\tau^{\prime}_{n}\tau_{n}},$$

$$\sum_{\tau_{n}\lambda_{n}}\left<\matrix{
[\lambda_{1n}] &[\lambda_{2n}]
\cr
[\lambda_{1~n-1}] &[\lambda_{2~n-1}]\cr}\right.\left\vert
\matrix{\tau_{n}~[\lambda_{n}]\cr
\tau_{n-1}~[\lambda_{n-1}]\cr}\right>
\left<\matrix{
[\lambda_{1n}] &[\lambda_{2n}]
\cr
[\lambda^{\prime}_{1~n-1}] &[\lambda^{\prime}_{2~n-1}]\cr}\right.\left\vert
\matrix{\tau_{n}~[\lambda_{n}]\cr
\tau_{n-1}~[\lambda_{n-1}]\cr}\right>$$

$$=\delta_{\lambda^{\prime}_{1~n-1}\lambda_{1~n-1}}
\delta_{\lambda^{\prime}_{2~n-1}\lambda_{2~n-1}}.\eqno(56)$$

\noindent In the following, we list no modification rule involved $SO(n)\supset SO(n-1)$ ISFs
for the coupling $[\lambda_{1}]\times [\lambda_{2}]$
with resulting irreps $[\lambda_{1n},\lambda_{2n},\lambda_{3n},\lambda_{4n},\dot{0}]$
for $\sum_{i}\lambda_{in}\leq 4$,
which are obtained from IDCs of
$S_{f_{1}}\times S_{f_{2}}\uparrow D_{f}(n)$ with $f_{1}+f_{2}=f\leq 4$. 

\vskip .7cm
\begin{centering}
{\large 6. Conclusions}\\
\end{centering}
\vskip .4cm
 In this paper, induced representations of $D_{f}(n)$ from $S_{f_{1}}\times S_{f_{2}}$ with
$f_{1}+f_{2}=f$ are constructed. The IDCs of  $S_{f_{1}}\times S_{f_{2}}\uparrow D_{f}(n)$ with $f\leq 4$ up
to a normalization factor are derived by using the linear equation method. It is found that
these IDCs are $SO(n)$ tensor component dependent.
Weyl tableaus for
the corresponding Gel'fand basis of SO(n) are defined. The assimilation method for obtaining
CGCs of SO(n) in the Gel'fand basis  with no modification rule involved 
couplings from IDCs of Brauer algebra are proposed,
which is based on the Brauer-Schur-Weyl duality relation between $O(n)$ and Brauer algebra
$D_{f}(n)$. 
Isoscalar factors of $SO(n)\supset SO(n-1)$ for the resulting irrep $[\lambda_{1},~\lambda_{2},~
\lambda_{3},~\lambda_{4},\dot{0}]$ with $\sum\limits_{i=1}^{4}\lambda_{i}\leq 4$ are tabulated.
 From these tables of ISFs, one can find that there are two types of no modification rule involved 
ISFs of $SO(n)\supset SO(n-1)$
 or CGCs of $SO(n)$ in its canonical basis. The type-one ISFs or CGCs are $n$-independent, which
are the same as ISFs of $U(n)\supset U(n-1)$ or CGCs of $U(n)$ in the canonical basis. Therefore,
The following type of ISFs of $U(n)\supset U(n-1)$ are also $SO(n)\supset
SO(n-1)$ ISFs:

$$\left<\matrix{
[\lambda_{1}] &[\lambda_{2}]
\cr
[\nu_{1}] &[\nu_{2}]\cr}\right.\left\vert
\matrix{\tau_{n}~~[\lambda]\cr
\tau_{n-1}[\nu]\cr}\right>\eqno(57)$$

\noindent with $\sum_{i}(\lambda_{1i}+\lambda_{2i})=\sum_{i}\lambda_{i}$,
$\sum_{j}(\nu_{1j}+\nu_{2j})=\sum_{j}\nu_{j}$, and $\sum_{i}\lambda_{i}-\sum_{j}\nu_{j}=0$ or $1$.
Hence, ISFs for $U(n)\supset U(n-1)$ of this type
or CGCs for $U(n)$ of this type derived previously,$^{[15-17]}$ in which many results are
with outer-multiplicity, are also those of $SO(n)$. From Brauer algebra point of view, 
there is no trace contraction between $[\lambda_{1}]$ and $[\lambda_{2}]$ in $SO(n)$
couplings in these cases. As a consequence, these coefficients can be derived from the IDCs of 
$S_{f_{1}}\times S_{f_{2}}\uparrow
S_{f}$, which has already been discussed in [15]. 
  The type-two ISFs of $SO(n)\supset SO(n-1)$ 
(CGCs of $SO(n)$ in the canonical basis) are rank-$n$ dependent, which can only be obtained
from IDCs of $S_{f_{1}}\times S_{f_{2}}\uparrow D_{f}(n)$. However, the problem concerning how to derive modification
rule involved  CGCs of $SO(n)$ from the IDCs of Brauer algebra still needs further study.

\vskip .8cm
\noindent {\bf Acknowledgment}
\vskip .5cm
This work was  supported by the US National Science Foundation through Grant
No. PHY-9603006 and Cooperative  Agreement No. EPS-9550481 which includes
matching from the Louisiana Board of Regents Support Fund, and was partly
supported by  National Natural Science Foundation of China.     

\vskip .5cm
\begin{tabbing}
\=1111\=22222222222222222222222222222222222222222222222222222222222222222222222222222222\=\kill\\
\>{[1]}\>{L. C. Biedenharn and J D Louck, Angular momentum in quantum Physics,}\\
\>{}\>{(Addison-Wesley Pub. Co., Reading, Mass., 1981)}.\\
\>{[2]}\>{L. C. Biedenharn, J. Math. Phys. {\bf 2}(1961) 433}\\
\>{[3]}\>{G. Racah, Phys. Rev. {\bf 76} (1949) 1352}.\\
\>{[4]}\>{S. J. Ali\v{s}auskas, J. Phys. A {\bf 20}(1987) 35}.\\
\>{[5]}\>{Feng Pan and Yu-Fang Cao, J. Math. Phys. {\bf 29}(1988) 2384}.\\
\>{[6]}\>{M. D. Gould, J. Math. Phys. {\bf 27} (1986) 1964}.\\
\>{[7]}\>{H. Z. Sun and D Ruan, J. Math. Phys. {\bf 39}(1998) 630}.\\
\>{[8]}\>{Feng Pan, J. Phys. A {\bf 28}(1995) 3139}.\\
\>{[9]}\>{R. Leduc and A. Ram, Adv. in Math. {\bf 125} (1997) 1}.\\
\>{[10]}\>{Feng Pan and L. Dai, J. Phys. A {\bf 29}(1996) 5079; 5093}.\\
\>{[11]}\>{Feng Pan, S. Dong, and J. P. Draayer, J. Phys. A 
{\bf 30}(1997) 8279}.\\
\>{[12]}\>{R. Brauer, Ann. Math. {\bf 63} (1937) 854}.\\
\>{[13]}\>{H. Wenzl, Ann. Math. {\bf 128} (1988) 173}.\\
\>{[14]}\>{Feng Pan and J. Q. Chen, J. Math. Phys. {\bf 34}(1993) 4305; 4316}.\\
\>{[15]}\>{J. Q. Chen,  P. N. Wang, Z. M. L\"{u}, and X. B. Wu, Tables of the CG, Racah, and}\\
\>{}\>{Subduction coefficients of $SU(n)$ groups, 
(Singapore, World Scientific, 1987)}.\\
\>{[16]}\>{S. J. Ali\v{s}auskas, A. A. Jucy, and A. P. Jucy, J. Math. Phys. 
{\bf 13}(1972) 1329 }.\\
\>{[17]}\>{J. P. Draayer and Y. Akiyama, J. Math. Phys. {\bf 14}(1973) 1904}.\\
\end{tabbing}

\newpage
\noindent Table 1. ISFs $\left<
\matrix{
[\lambda_{1}] &[\lambda_{2}]\cr
[\nu_{1}] &[\nu_{2}]\cr}\right\vert\left.
\matrix{
[\lambda ]\cr
[\nu ]\cr}\right>$ of $SO(n)\supset SO(n-1)$ for $[1]\times [1]=[2]+[1^{2}]^{*}+[0]$.
\begin{tabbing}
\=111111111111111111111111111111111111111111111111111111111111111111111111111111111111\=\kill\\
{------------------------------------------------------------------------------------------------------------------}\\
1111111111111111111\=22222222222222\=3333333333333\=4444444444444\=5555555555555\=6666666666666\=\kill\\
{$\matrix{
[\lambda ]\cr
[\nu ]\cr}$ /
$\matrix{ [\lambda_{1}] &[\lambda_{2}]\cr
[\nu_{1}] &[\nu_{2}]\cr}$} \>{$\matrix{ [10] &[10]\cr
[10] &[10]\cr}$}\>{$\matrix{ [10] &[10]\cr
[10] &[0]\cr}$}\>{$\matrix{ [10] &[10]\cr
[0] &[10]\cr}$}\>{$\matrix{ [10] &[10]\cr
[0] &[0]\cr}$}\\
\=111111111111111111111111111111111111111111111111111111111111111111111111111111111111\=\kill\\
{------------------------------------------------------------------------------------------------------------------}\\
1111111111111111111\=22222222222222\=3333333333333\=4444444444444\=5555555555555\=6666666666666\=\kill\\
{$\matrix{
[2]\cr
[2]\cr}$}\>{~~~1}\>{~~~0}\>{~~~0}\>{~~~~0}\\
{}\\
$\matrix{
[2]\cr
[1]\cr}$\>{~~~0}\>{$\sqrt{1\over{2}}$}\>{$\sqrt{1\over{2}}$}\>{~~~~~0}\\
{}\\
$\matrix{
[2]\cr
[0]\cr}$\>{$~~\sqrt{1\over{n}}$}\>{$~~~0$}\>{$~~~0$}\>{$-
\sqrt{n-1\over{n}}$}\\
{}\\
$\matrix{
[1^{2}]\cr
[1]\cr}$\>{~~~0}\>{$\sqrt{1\over{2}}$}\>{$-\sqrt{1\over{2}}$}\>{~~~~~0}\\
{}\\
{$\matrix{
[1^{2}]~~\cr
[1^{2}]^{**}\cr}$}\>{~~~1}\>{~~~0}\>{~~~0}\>{~~~~0}\\
{}\\
$\matrix{
[0]\cr
[0]\cr}$\>{$~~\sqrt{n-1\over{n}}$}\>{$~~~0$}\>{$~~~0$}\>{$\sqrt{1\over{n}}$}\\
\=111111111111111111111111111111111111111111111111111111111111111111111111111111111111\=\kill\\
{------------------------------------------------------------------------------------------------------------------}\\
\end{tabbing}
\noindent $^*$ The corresponding ISFs for this irrep are only valid for $n\geq 5$.\\
\noindent $^{**}$ It can be taken as [11] or [1-1] when $n=5$.

\newpage
\noindent Table 2. ISFs $\left<
\matrix{
[\lambda_{1}] &[\lambda_{2}]\cr
[\nu_{1}] &[\nu_{2}]\cr}\right\vert\left.
\matrix{
[\lambda ]\cr
[\nu ]\cr}\right>$ of $SO(n)\supset SO(n-1)$ for $[2]\times [1]=[30]+[21]^{*}+[1]$.
\begin{tabbing}
\=111111111111111111111111111111111111111111111111111111111111111111111111111111111111\=\kill\\
{-------------------------------------------------------------------------------------------------------------------}\\
11111111111111111\=222222222\=333333333\=4444444444\=5555555555\=666666666\=77777777\=88888888\=\kill\\
{$\matrix{
[\lambda ]\cr
[\nu ]\cr}$ /
$\matrix{ [\lambda_{1}] &[\lambda_{2}]\cr
[\nu_{1}] &[\nu_{2}]\cr}$} \>{$\matrix{ [2] &[1]\cr
[2] &[1]\cr}$}\>{$\matrix{ [2] &[1]\cr
[2] &[0]\cr}$}\>{$\matrix{ [2] &[1]\cr
[1] &[1]\cr}$}\>{$\matrix{ [2] &[1]\cr
[1] &[0]\cr}$}
\>{$\matrix{ [2] &[1]\cr
[0] &[1]\cr}$}
\>{$\matrix{ [2] &[1]\cr
[0] &[0]\cr}$}
\\
\=111111111111111111111111111111111111111111111111111111111111111111111111111111111111\=\kill\\
{-------------------------------------------------------------------------------------------------------------------}\\
11111111111111111\=222222222\=333333333\=4444444444\=5555555555\=666666666\=77777777\=88888888\=\kill\\
{$\matrix{
[3]\cr
[3]\cr}$}\>{~~~1}\>{~~~0}\>{~~~0}\>{~~~~0}\>{~~~0}\>{~~~~~0}\\
{}\\
$\matrix{
[3]\cr
[2]\cr}$\>{~~~0}\>{$\sqrt{1\over{3}}$}\>{$\sqrt{2\over{3}}$}\>{~~~~~0}\>{~~~~0}\>{~~~~~~0}\\
11111111111111\=222222222222\=333333333\=44444444\=5555555555\=666666666\=77777777\=88888888\=\kill\\
$\matrix{
[3]\cr
[1]\cr}$\>{$\sqrt{2(n-2)\over{3(n+2)(n-1)}}$}\>{$~~~0$}\>{$~~~0$}\>{$\sqrt{2(n+1)\over{3(n+2)}}$}\>{$-\sqrt{n(n+1)\over{3(n+2)(n-1)}}$}
\>{~~~~~~~~~~0}\\
{}\\
$\matrix{
[3]\cr
[0]\cr}$\>{~~~~~~~0}\>{~~~0}\>{$\sqrt{2\over{n+2}}$}\>{~~~~~~~0}\>{~~~~~~~0}\>{$~~~~~~\sqrt{n\over{n+2}}$}\\
{}\\
{$\matrix{
[21]~~\cr
[21]^{*2}\cr}$}\>{~~~~~~~1}\>{~~~~0}\>{~~~~0}\>{~~~~~~~~0}\>{~~~~~~~0}\>{~~~~~~~~~0}\\
{}\\
$\matrix{
[21]\cr
[2]\cr}$\>{~~~~~~~0}\>{$~~\sqrt{2\over{3}}$}\>{$-\sqrt{1\over{3}}$}\>{$~~~~~~~0$}\>{$~~~~~~~0$}\>{~~~~~~~~~0}\\
\>{}\\
$\matrix{
[21]\cr
[1]\cr}$\>{$\sqrt{n+1\over{3(n-1)^2}}$}\>{$~~~0$}\>{$~~~0$}\>{$~~\sqrt{n-2\over{3(n-1)}}$}\>{~~$\sqrt{2n(n-2)\over{3(n-1)^2}}$}
\>{~~~~~~~~~~0}\\
{}\\
{$\matrix{
[21]~~\cr
[1^2]^{*3}\cr}$}\>{~~~~~~~0}\>{~~~~0}\>{~~~~1}\>{~~~~~~~~0}\>{~~~~~~~0}\>{~~~~~~~~~~0}\\
11111111111111\=222222222222\=333333333\=444444\=5555555555555\=66666666666\=777777777\=88888888\=\kill\\
$\matrix{
[1]\cr
[1]\cr}$\>{$\sqrt{n(n+1)(n-2)\over{(n+2)(n-1)^2}}$}
\>{$~~~0$}\>{$~~~0$}\>{-$\sqrt{n\over{(n+2)(n-1)}}$}\>{-$\sqrt{2\over{(n+2)(n-1)^2}}$}
\>{~~~~~~0}\\
{}\\
$\matrix{
[1]\cr
[0]\cr}$\>{~~~~~~~0}\>{~~~0}\>{$\sqrt{n\over{n+2}}$}\>{~~~~~~~0}\>{~~~~~~~0}\>{~-$\sqrt{2\over{n+2}}$}\\
\=111111111111111111111111111111111111111111111111111111111111111111111111111111111111\=\kill\\
{-------------------------------------------------------------------------------------------------------------------}\\
\end{tabbing}
\noindent $^*$ The corresponding ISFs for this irrep are only valid for $n\geq 5$.~$^{*2}$ It can be taken as [21] or [2-1] when $n=5$.
$^{*3}$ It can be taken as [11] or [1-1] when $n=5$
\newpage
\noindent Table 3. ISFs $\left<
\matrix{
[\lambda_{1}] &[\lambda_{2}]\cr
[\nu_{1}] &[\nu_{2}]\cr}\right\vert\left.
\matrix{
[\lambda ]\cr
[\nu ]\cr}\right>$ of $SO(n)\supset SO(n-1)$ for $[30]\times [10]=[40]+[31]^{*}+[20]$.
\begin{tabbing}
\=111111111111111111111111111111111111111111111111111111111111111111111111111111111111\=\kill\\
{--------------------------------------------------------------------------------------------------------------------------------------}\\
1111111\=2222222222\=3333333333\=44444444444\=55555555555\=66666666666\=7777777777\=999999999\=\kill\\
{$\matrix{
[\lambda ]\cr
[\nu ]\cr}$ /}
\>{$\matrix{ [3] &[1]\cr
[3] &[1]\cr}$}\>{$\matrix{ [3] &[1]\cr
[3] &[0]\cr}$}\>{$\matrix{ [3] &[1]\cr
[2] &[1]\cr}$}\>{$\matrix{ [3] &[1]\cr
[2] &[0]\cr}$}
\>{$\matrix{ [3] &[1]\cr
[1] &[1]\cr}$}
\>{$\matrix{ [3] &[1]\cr
[1] &[0]\cr}$}\>{$\matrix{ [3] &[1]\cr
[0] &[1]\cr}$}\>{$\matrix{ [3] &[1]\cr
[0] &[0]\cr}$}\\
\=111111111111111111111111111111111111111111111111111111111111111111111111111111111111\=\kill\\
{--------------------------------------------------------------------------------------------------------------------------------------}\\
1111111\=22222222222\=333333333\=4444444444\=5555555555\=666666666\=77777777\=88888888\=\kill\\
{$\matrix{
[4]\cr
[4]\cr}$}\>{~~~~~~1}\>{~~~~0}\>{~~~~~~0}\>{~~~~~~~0}\>{~~~~~~~0}\>{~~~~~~~~~~0}\>{~~~~~~~~~~~~~~0}\>{~~~~~~~~~~~~~~0}\\
{}\\
$\matrix{
[4]\cr
[3]\cr}$\>{~~~~~0}\>{~~${1\over{2}}$}\>{~~$\sqrt{3\over{4}}$}\>{~~~~~~0}\>{~~~~~~~0}\>{~~~~~~~~~~0}\>{~~~~~~~~~~~~~~0}\>{~~~~~~~~~~~~~~0}\\
111111\=222222222222\=333333333\=44444444\=5555555555\=666666666\=77777777\=88888888\=\kill\\
$\matrix{
[4]\cr
[2]\cr}$\>{$\sqrt{n-1\over{2(n+1)(n+4)}}$}\>{$~~~0$}\>{$~~~~0$}\>{~-$\sqrt{n+3\over{2(n+4)}}$}\>{~~$\sqrt{(n+2)(n+3)\over{2(n+1)(n+4)}}$}
\>{~~~~~~~~~~~~~0}\>{~~~~~~~~~~~~~~~~~0}\>{~~~~~~~~~~~~~~~~~0}\\
1111111\=22222222222\=3333333\=4444444444\=5555555555\=666666666\=77777777\=88888888\=\kill\\
$\matrix{
[4]\cr
[1]\cr}$\>{~~~~~~~0}\>{~~~0}\>{$\sqrt{3(n-2)\over{2(n-1)(n+4)}}$}\>{~~~~~~~~~0}\>{~~~~~~~~~0}\>{$~~~~~~~~\sqrt{3(n+2)\over{4(n+4)}}$}
\>{~~~~~~~~~~$\sqrt{(n+1)(n+2)\over{4(n-1)(n+4)}}$}\>{~~~~~~~~~~~~~~~~~0}\\
{}\\
{$\matrix{
[4]\cr
[0]\cr}$}\>{~~~~~~~0}\>{~~~~0}\>{~~~~~~0}\>{~~~~~~~~0}\>{~~~~~~$\sqrt{3\over{n+4}}$}\>{~~~~~~~~~~~~~0}\>{~~~~~~~~~~~~~~~~~0}
\>{~~~~~~~~~~~~~$\sqrt{n+1\over{n+4}}$}\\
{}\\
$\matrix{
[31]\cr
[3]\cr}$\>{~~~~~~~0}\>{$~~\sqrt{3\over{4}}$}\>{~~~~~-${1\over{2}}$}\>{$~~~~~~~~0$}\>{$~~~~~~~~~0$}\>{~~~~~~~~~~~~0}
\>{~~~~~~~~~~~~~~~~0}\>{~~~~~~~~~~~~~~~~~0}\\
\=111111111111111111111111111111111111111111111111111111111111111111111111111111111111\=\kill\\
{---------------------------------------------------------------------------------------------------------------------------------------}\\
\end{tabbing}
\noindent $^{*}$ The corresponding ISFs for this irrep are only valid for $n\geq 5$.
\newpage
\noindent Table 3. (Continued)
\begin{tabbing}
\=111111111111111111111111111111111111111111111111111111111111111111111111111111111111\=\kill\\
{--------------------------------------------------------------------------------------------------------------------------------------}\\
1111111\=2222222222\=3333333333\=44444444444\=55555555555\=66666666666\=7777777777\=999999999\=\kill\\
{$\matrix{
[\lambda ]\cr
[\nu ]\cr}$ /}
\>{$\matrix{ [3] &[1]\cr
[3] &[1]\cr}$}\>{$\matrix{ [3] &[1]\cr
[3] &[0]\cr}$}\>{$\matrix{ [3] &[1]\cr
[2] &[1]\cr}$}\>{$\matrix{ [3] &[1]\cr
[2] &[0]\cr}$}
\>{$\matrix{ [3] &[1]\cr
[1] &[1]\cr}$}
\>{$\matrix{ [3] &[1]\cr
[1] &[0]\cr}$}\>{$\matrix{ [3] &[1]\cr
[0] &[1]\cr}$}\>{$\matrix{ [3] &[1]\cr
[0] &[0]\cr}$}\\
\=111111111111111111111111111111111111111111111111111111111111111111111111111111111111\=\kill\\
{--------------------------------------------------------------------------------------------------------------------------------------}\\
1111111\=22222222222\=333333333\=4444444444\=5555555555\=666666666\=77777777\=88888888\=\kill\\
$\matrix{
[31]\cr
[2]\cr}$\>{$\sqrt{n+3\over{2n(n+1)}}$}\>{$~~~0$}\>{$~~~0$}\>{~~~-$\sqrt{n-1\over{2n}}$}\>{~-$\sqrt{(n-1)(n+2)\over{2n(n+1)}}$}
\>{~~~~~~~~~~~~0}\>{~~~~~~~~~~~~~~~~0}\>{~~~~~~~~~~~~~~~~~0}\\
{}\\
$\matrix{
[31]\cr
[1]\cr}$\>{~~~~~~$0$}\>{$~~~0$}\>{$\sqrt{n+2\over{2n(n-1)}}$}\>{~~~~~~~$0$}\>{~~~~~~$~~~0$}
\>{~~~~~~~~$\sqrt{n-2\over{4n}}$}\>{~~~~~~~-$\sqrt{3(n+1)(n-2)\over{4n(n-1)}}$}\>{~~~~~~~~~~~~~~~~~0}\\
\>{}\\
$\matrix{
[31]\cr
[21]^{*}\cr}$\>{~~~~~~$0$}\>{$~~~0$}\>{$~~~1$}\>{~~~~~~~~0}\>{~~~~~~~~~$0$}
\>{~~~~~~~~~~~~0}\>{~~~~~~~~~~~~~~~~0}\>{~~~~~~~~~~~~~~~~~0}\\
{}\\
$\matrix{
[31]~~\cr
[1^{2}]^{**}\cr}$\>{~~~~~~$0$}\>{$~~~0$}\>{$~~~0$}\>{~~~~~~~~0}\>{~~~~~~~~~$1$}
\>{~~~~~~~~~~~~0}\>{~~~~~~~~~~~~~~~~0}\>{~~~~~~~~~~~~~~~~~0}\\
1111\=22222222222222\=333333333333\=4444444444\=5555555555\=666666666\=77777777\=88888888\=\kill\\
$\matrix{
[2]\cr
[2]\cr}$\>{$\sqrt{(n+2)(n+3)(n-1)\over{n(n+1)(n+4)}}$}\>{$~~~0$}\>{$0$}\>{$\sqrt{n+2\over{n(n+4)}}$}
\>{$\sqrt{4\over{n(n+1)(n+4)}}$}
\>{~~~~~~~~~0}\>{~~~~~~~~~~~~~0}\>{~~~~~~~~~~~~~0}\\
1111\=22222222222222\=3333333\=4444444444\=5555555555\=666666666\=77777777\=88888888\=\kill\\
$\matrix{
[2]\cr
[1]\cr}$\>{$~~~~~~~~~~0$}\>{$~~~0$}\>{$\sqrt{(n^2-4)(n+1)\over{n(n-1)(n+4)}}$}\>{~~~~~~~~~~~$0$}
\>{~~~~~~~~~~~~0}\>{~~~~~~~~~-$\sqrt{2(n+1)\over{n(n+4)}}$}\>{~~~~~~~~~~~~$\sqrt{6\over{n(n-1)(n+4)}}$}
\>{~~~~~~~~~~~~~~~~~~~~~0}\\
{}\\
$\matrix{
[2]\cr
[0]\cr}$\>{~~~~~~~~~~$0$}\>{$~~~0$}\>{$~~~~~~~~~0$}\>{~~~~~~~~~$~~0$}\>{~~~~~~~~~$\sqrt{n+1\over{n+4}}$}
\>{~~~~~~~~~~~~~~~~0}\>{~~~~~~~~~~~~~~~~~~~~0}\>{~~~~~~~~~~~~~~~~-$\sqrt{3\over{n+4}}$}\\

\=111111111111111111111111111111111111111111111111111111111111111111111111111111111111\=\kill\\
{--------------------------------------------------------------------------------------------------------------------------------------}\\
\end{tabbing}
\noindent $^{*}$ It can be taken as [21] or [2-1] when $n=5$.
$^{**}$ It can be taken as [11] or [1-1] when $n=5$
\newpage
\noindent Table 4. ISFs $\left<
\matrix{
[\lambda_{1}] &[\lambda_{2}]\cr
[\nu_{1}] &[\nu_{2}]\cr}\right\vert\left.
\matrix{
[\lambda ]\cr
[\nu ]\cr}\right>$ of $SO(n)\supset SO(n-1)$ for $[1^{2}]\times [1]=[21]+[1^{3}]+[1]$ for $n\geq 7$.
\begin{tabbing}
\=111111111111111111111111111111111111111111111111111111111111111111111111111111111111\=\kill\\
{------------------------------------------------------------------------------------------------------------------}\\
1111111111111111111\=22222222222222\=3333333333333\=4444444444444\=5555555555555\=6666666666666\=\kill\\
{$\matrix{
[\lambda ]\cr
[\nu ]\cr}$ /
$\matrix{ [\lambda_{1}] &[\lambda_{2}]\cr
[\nu_{1}] &[\nu_{2}]\cr}$} \>{$\matrix{ [1^{2}] &[10]\cr
[1^2] &[10]\cr}$}\>{$\matrix{ [1^2] &[10]\cr
[1^2] &[0]\cr}$}\>{$\matrix{ [1^2] &[10]\cr
[1] &[1]\cr}$}\>{$\matrix{ [1^2] &[10]\cr
[1] &[0]\cr}$}\\
\=111111111111111111111111111111111111111111111111111111111111111111111111111111111111\=\kill\\
{------------------------------------------------------------------------------------------------------------------}\\
1111111111111111111\=22222222222222\=3333333333333\=4444444444444\=5555555555555\=6666666666666\=\kill\\
{$\matrix{
[21]\cr
[21]\cr}$}\>{~~~1}\>{~~~0}\>{~~~0}\>{~~~~0}\\
{}\\
{$\matrix{
[21]\cr
[2]\cr}$}\>{~~~0}\>{~~~0}\>{~~~1}\>{~~~~0}\\
{}\\
$\matrix{
[21]\cr
[1^2]\cr}$\>{~~~0}\>{$\sqrt{2\over{3}}$}\>{$\sqrt{1\over{3}}$}\>{~~~~~0}\\
{}\\
$\matrix{
[21]\cr
[1]\cr}$\>{$\sqrt{1\over{n-1}}$}\>{$~~~0$}\>{$~~~0$}\>{$\sqrt{n-2\over{n-1}}$}\\
{}\\
{$\matrix{
[1^3]~\cr
[1^3]^{*}\cr}$}\>{~~~1}\>{~~~0}\>{~~~0}\>{~~~~0}\\
{}\\
$\matrix{
[1^{3}]\cr
[1]\cr}$\>{~~~0}\>{$\sqrt{1\over{3}}$}\>{$-\sqrt{2\over{3}}$}\>{~~~~~0}\\
{}\\
$\matrix{
[1]\cr
[1]\cr}$\>{$\sqrt{n-2\over{n-1}}$}\>{$~~~0$}\>{$~~~0$}\>{-$\sqrt{1\over{n-1}}$}\\
{}\\
{$\matrix{
[1]\cr
[0]\cr}$}\>{~~~0}\>{~~~0}\>{~~~1}\>{~~~~0}\\
\=111111111111111111111111111111111111111111111111111111111111111111111111111111111111\=\kill\\
{------------------------------------------------------------------------------------------------------------------}\\
\end{tabbing}
\noindent $^{*}$ It can be taken as [111] or [11-1] when $n=7$
\newpage
\noindent Table 5. ISFs $\left<
\matrix{
[\lambda_{1}] &[\lambda_{2}]\cr
[\nu_{1}] &[\nu_{2}]\cr}\right\vert\left.
\matrix{
[\lambda ]\cr
[\nu ]\cr}\right>$ of $SO(n)\supset SO(n-1)$ for $[1^{3}]\times [1]=[211]+[1^{4}]+[1^2]$ for $n\geq 9$.
\begin{tabbing}
\=111111111111111111111111111111111111111111111111111111111111111111111111111111111111\=\kill\\
{------------------------------------------------------------------------------------------------------------------}\\
1111111111111111111\=22222222222222\=3333333333333\=4444444444444\=5555555555555\=6666666666666\=\kill\\
{$\matrix{
[\lambda ]\cr
[\nu ]\cr}$ /
$\matrix{ [\lambda_{1}] &[\lambda_{2}]\cr
[\nu_{1}] &[\nu_{2}]\cr}$} \>{$\matrix{ [1^{3}] &[10]\cr
[1^3] &[10]\cr}$}\>{$\matrix{ [1^3] &[10]\cr
[1^3] &[0]\cr}$}\>{$\matrix{ [1^3] &[10]\cr
[1^2] &[1]\cr}$}\>{$\matrix{ [1^3] &[10]\cr
[1^2] &[0]\cr}$}\\
\=111111111111111111111111111111111111111111111111111111111111111111111111111111111111\=\kill\\
{------------------------------------------------------------------------------------------------------------------}\\
1111111111111111111\=22222222222222\=3333333333333\=4444444444444\=5555555555555\=6666666666666\=\kill\\
{$\matrix{
[211]\cr
[211]\cr}$}\>{~~~1}\>{~~~0}\>{~~~0}\>{~~~~0}\\
{}\\
{$\matrix{
[211]\cr
[21]\cr}$}\>{~~~0}\>{~~~0}\>{~~~1}\>{~~~~0}\\
{}\\
$\matrix{
[211]\cr
[1^2]\cr}$\>{$\sqrt{1\over{n-2}}$}\>{$~~~0$}\>{$~~~0$}\>{$\sqrt{n-3\over{n-2}}$}\\
{}\\
$\matrix{
[211]\cr
[1^3]\cr}$\>{~~~0}\>{$\sqrt{3\over{4}}$}\>{~~${1\over{2}}$}\>{~~~~~0}\\
{}\\
{$\matrix{
[1^4]~\cr
[1^4]^{*}\cr}$}\>{~~~1}\>{~~~0}\>{~~~0}\>{~~~~0}\\
{}\\
$\matrix{
[1^{4}]\cr
[1^3]\cr}$\>{~~~0}\>{~~${1\over{2}}$}\>{$-\sqrt{3\over{4}}$}\>{~~~~~0}\\
{}\\
$\matrix{
[1^2]\cr
[1^2]\cr}$\>{$\sqrt{n-3\over{n-2}}$}\>{$~~~0$}\>{$~~~0$}\>{-$\sqrt{1\over{n-2}}$}\\
{}\\
{$\matrix{
[1^2]\cr
[1]\cr}$}\>{~~~0}\>{~~~0}\>{~~~1}\>{~~~~0}\\
\=111111111111111111111111111111111111111111111111111111111111111111111111111111111111\=\kill\\
{------------------------------------------------------------------------------------------------------------------}\\
\end{tabbing}
\noindent $^{*}$ It can be taken as [1111] or [111-1] when $n=9$
\newpage
\noindent Table 6. ISFs $\left<
\matrix{
[\lambda_{1}] &[\lambda_{2}]\cr
[\nu_{1}] &[\nu_{2}]\cr}\right\vert\left.
\matrix{
[\lambda ]\cr
[\nu ]\cr}\right>$ of $SO(n)\supset SO(n-1)$ for $[2]\times [1^2]=[31]+[211]+[2]+[1^2]$ for $n\geq 7$.
\begin{tabbing}
\=111111111111111111111111111111111111111111111111111111111111111111111111111111111111\=\kill\\
{------------------------------------------------------------------------------------------------------------------}\\
11111111111111111\=222222222\=333333333\=4444444444\=5555555555\=666666666\=77777777\=88888888\=\kill\\
{$\matrix{
[\lambda ]\cr
[\nu ]\cr}$ /
$\matrix{ [\lambda_{1}] &[\lambda_{2}]\cr
[\nu_{1}] &[\nu_{2}]\cr}$} \>{$\matrix{ [2] &[1^2]\cr
[2] &[1^2]\cr}$}\>{$\matrix{ [2] &[1^2]\cr
[2] &[1]\cr}$}\>{$\matrix{ [2] &[1^2]\cr
[1] &[1^2]\cr}$}\>{$\matrix{ [2] &[1^2]\cr
[1] &[1]\cr}$}
\>{$\matrix{ [2] &[1^2]\cr
[0] &[1^2]\cr}$}
\>{$\matrix{ [2] &[1^2]\cr
[0] &[1]\cr}$}
\\
\=111111111111111111111111111111111111111111111111111111111111111111111111111111111111\=\kill\\
{------------------------------------------------------------------------------------------------------------------}\\
11111111111111111\=222222222\=333333333\=4444444444\=5555555555\=666666666\=77777777\=88888888\=\kill\\
{$\matrix{
[31]\cr
[31]\cr}$}\>{~~~1}\>{~~~0}\>{~~~0}\>{~~~~0}\>{~~~0}\>{~~~~~0}\\
{}\\
{$\matrix{
[31]\cr
[3]\cr}$}\>{~~~0}\>{~~~1}\>{~~~0}\>{~~~~0}\>{~~~0}\>{~~~~~0}\\
{}\\
$\matrix{
[31]\cr
[21]\cr}$\>{~~~0}\>{~~-${1\over{2}}$}\>{$\sqrt{3\over{4}}$}\>{~~~~~0}\>{~~~~0}\>{~~~~~~0}\\
{}\\
{$\matrix{
[31]\cr
[2]\cr}$}\>{~$\sqrt{1\over{n}}$}\>{~~~0}\>{~~~0}\>{$\sqrt{n-1\over{n}}$}\>{~~~0}\>{~~~~~~0}\\
{}\\
11111111111111\=222222222222\=333333333\=44444444\=5555555555\=666666666\=77777777\=88888888\=\kill\\
$\matrix{
[31]\cr
[1^2]\cr}$\>{$\sqrt{n-3\over{2(n+2)(n-1)}}$}\>{$~~~0$}\>{$~~~0$}\>{$\sqrt{n+1\over{2(n+2)}}$}\>{$-\sqrt{n(n+1)\over{2(n+2)(n-1)}}$}
\>{~~~~~~~~~~0}\\
{}\\
1111111111111\=2222222\=3333333333333\=4444444444\=5555555555\=666666666\=77777777\=88888888\=\kill\\
$\matrix{
[31]\cr
[1]\cr}$\>{~~~~~~~0}\>{$\sqrt{2(n+2)(n-1)\over{n(n+1)^2(n-2)^2}}$}
\>{$\sqrt{2(n-1)^2(n+2)\over{n(n+1)(n-2)^2}}$}\>{~~~~~~~0}
\>{$~~\sqrt{(n-1)(n+2)\over{(n+1)(n-2)}}$}
\>{~~~~~~~0}\\
{}\\
{$\matrix{
[211]~\cr
[211]^{*}\cr}$}\>{~~~~~~~1}\>{~~~~~~~~~~~~0}\>{~~~~~~0}\>{~~~~~~~~0}\>{~~~~~~~0}\>{~~~~~~~~~0}\\
\=111111111111111111111111111111111111111111111111111111111111111111111111111111111111\=\kill\\
{------------------------------------------------------------------------------------------------------------------}\\
\end{tabbing}
\noindent $^{*}$ It can be taken as [211] or [21-1] when $n=7$
\newpage
\noindent Table 6. (Continued)
\begin{tabbing}
\=111111111111111111111111111111111111111111111111111111111111111111111111111111111111\=\kill\\
{------------------------------------------------------------------------------------------------------------------}\\
11111111111111111\=222222222\=333333333\=4444444444\=5555555555\=666666666\=77777777\=88888888\=\kill\\
{$\matrix{
[\lambda ]\cr
[\nu ]\cr}$ /
$\matrix{ [\lambda_{1}] &[\lambda_{2}]\cr
[\nu_{1}] &[\nu_{2}]\cr}$} \>{$\matrix{ [2] &[1^2]\cr
[2] &[1^2]\cr}$}\>{$\matrix{ [2] &[1^2]\cr
[2] &[1]\cr}$}\>{$\matrix{ [2] &[1^2]\cr
[1] &[1^2]\cr}$}\>{$\matrix{ [2] &[1^2]\cr
[1] &[1]\cr}$}
\>{$\matrix{ [2] &[1^2]\cr
[0] &[1^2]\cr}$}
\>{$\matrix{ [2] &[1^2]\cr
[0] &[1]\cr}$}
\\
\=111111111111111111111111111111111111111111111111111111111111111111111111111111111111\=\kill\\
{------------------------------------------------------------------------------------------------------------------}\\
11111111111111111\=222222222\=333333333\=4444444444\=5555555555\=666666666\=77777777\=88888888\=\kill\\
$\matrix{
[211]\cr
[21]\cr}$\>{~~~~~~~0}\>{$~~\sqrt{3\over{4}}$}\>{~~~~~${1\over{2}}$}\>{$~~~~~~~0$}\>{$~~~~~~~0$}\>{~~~~~~~~0}\\
\>{}\\
{$\matrix{
[211]~\cr
[1^3]^{*}\cr}$}\>{~~~~~~~0}\>{~~~~0}\>{~~~~~1}\>{~~~~~~~0}\>{~~~~~~0}\>{~~~~~~~~0}\\
11111111111111\=222222222222\=333333333\=44444444\=5555555555\=666666666\=77777777\=88888888\=\kill\\
$\matrix{
[211]\cr
[1^2]\cr}$\>{$\sqrt{n+1\over{2(n-2)(n-1)}}$}\>{$~~~~0$}\>{$~~~~0$}\>{~~$\sqrt{n-3\over{2(n-2)}}$}\>{$-\sqrt{n(n-3)\over{2(n-2)(n-1)}}$}
\>{~~~~~~~~~~0}\\
{}\\
{$\matrix{
[2]\cr
[2]\cr}$}\>{~~~~~$\sqrt{n-1\over{n}}$}\>{~~~~0}\>{~~~~0}\>{~~~~~~$\sqrt{1\over{n}}$}\>{~~~~~~~~~0}\>{~~~~~~~~~~0}\\
1111111111111111\=222222222\=333333333\=4444444444\=5555555555\=666666666\=77777777\=88888888\=\kill\\
{$\matrix{
[2]\cr
[1]\cr}$}\>{~~~~~~~~0}\>{$\sqrt{(n-2)(n+1)\over{2n(n-1)}}$}\>{~~~$\sqrt{n-2\over{2n}}$}\>{~~~~~~~~0}\>{~~~~~~~~~0}
\>{~~~-$\sqrt{1\over{n-1}}$}\\
\>{}\\
{$\matrix{
[2]\cr
[0]\cr}$}\>{~~~~~~~0}\>{~~~~~~0}\>{~~~~~0}\>{~~~~~~~1}\>{~~~~~~~~0}\>{~~~~~~~~0}\\
11111111111111\=222222222222\=333333333\=444444\=5555555555555\=66666666666\=777777777\=88888888\=\kill\\
$\matrix{
[1^2]\cr
[1^2]\cr}$\>{$\sqrt{n(n+1)(n-3)\over{(n-1)(n^2-4)}}$}
\>{$~~~0$}\>{$~~~0$}\>{-$\sqrt{n\over{(n+2)(n-2)}}$}\>{$\sqrt{4\over{(n^2-4)(n-1)}}$}
\>{~~~~~~0}\\
11111111111111\=22222222\=333333333333\=444444444\=5555555555555\=66666666\=77777\=88888\=\kill\\
$\matrix{
[1^2]\cr
[1]\cr}$\>{~~~~~~~~0}\>{$\sqrt{n(n+1)\over{2(n+2)(n-1)}}$}\>{-$\sqrt{n\over{2(n+2)}}$}\>{~~~~~~~0}\>{~~~~~0}
\>{$\sqrt{n-2\over{(n+2)(n-1)}}$}\\
\=111111111111111111111111111111111111111111111111111111111111111111111111111111111111\=\kill\\
{------------------------------------------------------------------------------------------------------------------}\\
\end{tabbing}
\noindent $^{*}$ It can be taken as [111] or [11-1] when $n=7$
\newpage
\noindent Table 7. ISFs $\left<
\matrix{
[\lambda_{1}] &[\lambda_{2}]\cr
[\nu_{1}] &[\nu_{2}]\cr}\right\vert\left.
\matrix{
[\lambda ]\cr
[\nu ]\cr}\right>$ of $SO(n)\supset SO(n-1)$ for $[1^2]\times [1^2]=[22]+[21^{2}]$\\
\vskip .2cm
\noindent $+[1^4]^{*}+[20]+[1^2]+[0]$ for $n\geq 7$.
\begin{tabbing}
\=111111111111111111111111111111111111111111111111111111111111111111111111111111111111\=\kill\\
{------------------------------------------------------------------------------------------------------------------}\\
1111111111111111111\=22222222222222\=3333333333333\=4444444444444\=5555555555555\=6666666666666\=\kill\\
{$\matrix{
[\lambda ]\cr
[\nu ]\cr}$ /
$\matrix{ [\lambda_{1}] &[\lambda_{2}]\cr
[\nu_{1}] &[\nu_{2}]\cr}$} \>{$\matrix{ [1^2] &[1^2]\cr
[1^2] &[1^2]\cr}$}\>{$\matrix{ [1^2] &[1^2]\cr
[1^2] &[1]\cr}$}\>{$\matrix{ [1^2] &[1^2]\cr
[1] &[1^2]\cr}$}\>{$\matrix{ [1^2] &[1^2]\cr
[1] &[1]\cr}$}\\
\=111111111111111111111111111111111111111111111111111111111111111111111111111111111111\=\kill\\
{------------------------------------------------------------------------------------------------------------------}\\
1111111111111111111\=22222222222222\=3333333333333\=4444444444444\=5555555555555\=6666666666666\=\kill\\
{$\matrix{
[22]\cr
[22]\cr}$}\>{~~~1}\>{~~~0}\>{~~~0}\>{~~~~0}\\
{}\\
$\matrix{
[22]\cr
[21]\cr}$\>{~~~0}\>{$\sqrt{1\over{2}}$}\>{$\sqrt{1\over{2}}$}\>{~~~~~0}\\
{}\\
$\matrix{
[22]\cr
[2]\cr}$\>{$~~\sqrt{1\over{n-2}}$}\>{$~~~0$}\>{$~~~0$}\>{$\sqrt{n-3\over{n-2}}$}\\
{}\\
{$\matrix{
[211]~~\cr
[211]^{*2}\cr}$}\>{~~~1}\>{~~~0}\>{~~~0}\>{~~~~0}\\
{}\\
$\matrix{
[21^{2}]\cr
[21]\cr}$\>{~~~0}\>{$\sqrt{1\over{2}}$}\>{$-\sqrt{1\over{2}}$}\>{~~~~~0}\\
{}\\
$\matrix{
[21^{2}]\cr
[1^2]\cr}$\>{~~~0}\>{-$\sqrt{1\over{2}}$}\>{-$\sqrt{1\over{2}}$}\>{~~~~~0}\\
{}\\
$\matrix{
[21^2]\cr
[1^2]\cr}$\>{$~~\sqrt{1\over{n-2}}$}\>{$~~~0$}\>{$~~~0$}\>{$\sqrt{n-3\over{n-2}}$}\\
{}\\
{$\matrix{
[1^{4}]~~\cr
[1^{4}]^{*3}\cr}$}\>{~~~1}\>{~~~0}\>{~~~0}\>{~~~~0}\\
\=111111111111111111111111111111111111111111111111111111111111111111111111111111111111\=\kill\\
{------------------------------------------------------------------------------------------------------------------}\\
\end{tabbing}
\noindent $^*$ The corresponding ISFs for this irrep are only valid for $n\geq 9$.
\noindent $^{*2}$ It can be taken as [211] or [21-1] when $n=7$.
$^{*3}$ It can be taken as [1111] or [111-1] when $n=9$
\newpage
\noindent Table 7. (Continued)
\begin{tabbing}
\=111111111111111111111111111111111111111111111111111111111111111111111111111111111111\=\kill\\
{------------------------------------------------------------------------------------------------------------------}\\
1111111111111111111\=22222222222222\=3333333333333\=4444444444444\=5555555555555\=6666666666666\=\kill\\
{$\matrix{
[\lambda ]\cr
[\nu ]\cr}$ /
$\matrix{ [\lambda_{1}] &[\lambda_{2}]\cr
[\nu_{1}] &[\nu_{2}]\cr}$} \>{$\matrix{ [1^2] &[1^2]\cr
[1^2] &[1^2]\cr}$}\>{$\matrix{ [1^2] &[1^2]\cr
[1^2] &[1]\cr}$}\>{$\matrix{ [1^2] &[1^2]\cr
[1] &[1^2]\cr}$}\>{$\matrix{ [1^2] &[1^2]\cr
[1] &[1]\cr}$}\\
\=111111111111111111111111111111111111111111111111111111111111111111111111111111111111\=\kill\\
{------------------------------------------------------------------------------------------------------------------}\\
1111111111111111111\=22222222222222\=3333333333333\=4444444444444\=5555555555555\=6666666666666\=\kill\\
$\matrix{
[1^4]\cr
[1^3]\cr}$\>{~~~0}\>{$\sqrt{1\over{2}}$}\>{-$\sqrt{1\over{2}}$}\>{~~~~~0}\\
{}\\
$\matrix{
[2]\cr
[2]\cr}$\>{$~~\sqrt{n-3\over{n-2}}$}\>{$~~~0$}\>{$~~~0$}\>{-$\sqrt{1\over{n-2}}$}\\
{}\\
$\matrix{
[2]\cr
[1]\cr}$\>{~~~0}\>{$\sqrt{1\over{2}}$}\>{$\sqrt{1\over{2}}$}\>{~~~~~0}\\
{}\\
$\matrix{
[2]\cr
[0]\cr}$\>{$~~\sqrt{2\over{n}}$}\>{$~~~0$}\>{$~~~0$}\>{$\sqrt{n-2\over{n}}$}\\
{}\\
$\matrix{
[1^2]\cr
[1^2]\cr}$\>{$~~\sqrt{n-3\over{n-2}}$}\>{$~~~0$}\>{$~~~0$}\>{-$\sqrt{1\over{n-2}}$}\\
{}\\
$\matrix{
[1^2]\cr
[1]\cr}$\>{~~~0}\>{$\sqrt{1\over{2}}$}\>{-$\sqrt{1\over{2}}$}\>{~~~~~0}\\
{}\\
{$\matrix{
[0]\cr
[0]\cr}$}\>{~~$\sqrt{n-2\over{n}}$}\>{~~~0}\>{~~~0}\>{-$\sqrt{2\over{n}}$}\\
\=111111111111111111111111111111111111111111111111111111111111111111111111111111111111\=\kill\\
{------------------------------------------------------------------------------------------------------------------}\\
\end{tabbing}
\newpage
\noindent Table 8. ISFs $\left<
\matrix{
[\lambda_{1}] &[\lambda_{2}]\cr
[\nu_{1}] &[\nu_{2}]\cr}\right\vert\left.
\matrix{
[\lambda ]\cr
[\nu ]\cr}\right>$ of $SO(n)\supset SO(n-1)$ for $[21]\times [10]=[31]+[22]+$\\
\vskip .2cm
\noindent $[211]+[20]+[1^2]$ for $n\geq 7$.
\begin{tabbing}
\=111111111111111111111111111111111111111111111111111111111111111111111111111111111111\=\kill\\
{--------------------------------------------------------------------------------------------------------------------------------------}\\
1111111\=2222222222\=3333333333\=44444444444\=55555555555\=66666666666\=7777777777\=999999999\=\kill\\
{$\matrix{
[\lambda ]\cr
[\nu ]\cr}$ /}
\>{$\matrix{ [21] &[1]\cr
[21] &[1]\cr}$}\>{$\matrix{ [21] &[1]\cr
[21] &[0]\cr}$}\>{$\matrix{ [21] &[1]\cr
[2] &[1]\cr}$}\>{$\matrix{ [21] &[1]\cr
[2] &[0]\cr}$}
\>{$\matrix{ [21] &[1]\cr
[1^2] &[1]\cr}$}
\>{$\matrix{ [21] &[1]\cr
[1^2] &[0]\cr}$}\>{$\matrix{ [21] &[1]\cr
[1] &[1]\cr}$}\>{$\matrix{ [21] &[1]\cr
[1] &[0]\cr}$}\\
\=111111111111111111111111111111111111111111111111111111111111111111111111111111111111\=\kill\\
{--------------------------------------------------------------------------------------------------------------------------------------}\\
1111111\=22222222222\=333333333\=4444444444\=5555555555\=666666666\=77777777\=88888888\=\kill\\
{$\matrix{
[31]\cr
[31]\cr}$}\>{~~~~~~1}\>{~~~~0}\>{~~~~~~0}\>{~~~~~~~0}\>{~~~~~~~0}\>{~~~~~~~~~~0}\>{~~~~~~~~~~~~~~0}\>{~~~~~~~~~~~~~~0}\\
{}\\
{$\matrix{
[22]\cr
[22]\cr}$}\>{~~~~~~1}\>{~~~~0}\>{~~~~~~0}\>{~~~~~~~0}\>{~~~~~~~0}\>{~~~~~~~~~~0}\>{~~~~~~~~~~~~~~0}\>{~~~~~~~~~~~~~~0}\\
{}\\
{$\matrix{
[31]\cr
[3]\cr}$}\>{~~~~~~0}\>{~~~~0}\>{~~~~~~1}\>{~~~~~~~0}\>{~~~~~~~0}\>{~~~~~~~~~~0}\>{~~~~~~~~~~~~~~0}\>{~~~~~~~~~~~~~~0}\\
{}\\
$\matrix{
[31]\cr
[21]\cr}$\>{~~~~~~0}\>{~~$\sqrt{3\over{8}}$}\>{~~~~~~${1\over{4}}$}\>{~~~~~~0}\>{~~~~~~$3\over{4}$}\>{~~~~~~~~~~0}\>{~~~~~~~~~~~~~~0}\>{~~~~~~~~~~~~~~0}\\
111111\=222222222222\=333333333\=44444444\=5555555555\=666666666\=77777777\=88888888\=\kill\\
$\matrix{
[31]\cr
[2]\cr}$\>{~~~$\sqrt{n-3\over{2n(n-2)}}$}\>{$~~~~0$}\>{$~~~~~0$}\>{~~~~$\sqrt{n-1\over{2n}}$}\>{~~~~~$\sqrt{(n-1)^2\over{2n(n-2)}}$}
\>{~~~~~~~~~~~~~0}\>{~~~~~~~~~~~~~~~~~0}\>{~~~~~~~~~~~~~~~~~0}\\
{}\\
$\matrix{
[31]\cr
[1^2]\cr}$\>{~~~$\sqrt{3(n-3)\over{4(n^2-4)}}$}\>{~~~~0}\>{~~~~~~0}\>{~~~~~~~~~0}\>{~~~~~~~~0}\>{~~~~~-$\sqrt{3(n+1)\over{4(n+2)}}$}
\>{~~~~~~~~~~-$\sqrt{n^2-1\over{4(n^2-4)}}$}
\>{~~~~~~~~~~~~~~~~~0}\\
{}\\
{$\matrix{
[31]\cr
[1]\cr}$}\>{~~~~~~~0}\>{~~~~0}\>{$\sqrt{n-1\over{2n(n+2)}}$}\>{~~~~~~~~0}\>{~~~-$\sqrt{3(n+1)\over{2n(n+2)}}$}\>{~~~~~~~~~~~~~0}\>{~~~~~~~~~~~~~~~~~0}
\>{~~~~~~~~~~-$\sqrt{n^2-1\over{n(n+2)}}$}\\
{}\\
$\matrix{
[22]\cr
[21]\cr}$\>{~~~~~~~0}\>{$~~~{1\over{2}}$}\>{~~~$\sqrt{3\over{8}}$}\>{$~~~~~~~~0$}\>{~~~~~~-$\sqrt{3\over{8}}$}\>{~~~~~~~~~~~~0}
\>{~~~~~~~~~~~~~~~~0}\>{~~~~~~~~~~~~~~~~~0}\\
\=111111111111111111111111111111111111111111111111111111111111111111111111111111111111\=\kill\\
{---------------------------------------------------------------------------------------------------------------------------------------}\\
\end{tabbing}
\newpage
\noindent Table 8. (Continued)
\begin{tabbing}
\=111111111111111111111111111111111111111111111111111111111111111111111111111111111111\=\kill\\
{--------------------------------------------------------------------------------------------------------------------------------------}\\
1111111\=2222222222\=3333333333\=44444444444\=55555555555\=66666666666\=7777777777\=999999999\=\kill\\
{$\matrix{
[\lambda ]\cr
[\nu ]\cr}$ /}
\>{$\matrix{ [21] &[1]\cr
[21] &[1]\cr}$}\>{$\matrix{ [21] &[1]\cr
[21] &[0]\cr}$}\>{$\matrix{ [21] &[1]\cr
[2] &[1]\cr}$}\>{$\matrix{ [21] &[1]\cr
[2] &[0]\cr}$}
\>{$\matrix{ [21] &[1]\cr
[1^2] &[1]\cr}$}
\>{$\matrix{ [21] &[1]\cr
[1^2] &[0]\cr}$}\>{$\matrix{ [21] &[1]\cr
[1] &[1]\cr}$}\>{$\matrix{ [21] &[1]\cr
[1] &[0]\cr}$}\\
\=111111111111111111111111111111111111111111111111111111111111111111111111111111111111\=\kill\\
{--------------------------------------------------------------------------------------------------------------------------------------}\\
1111111\=22222222222\=333333333\=4444444444\=5555555555\=666666666\=77777777\=88888888\=\kill\\
$\matrix{
[22]\cr
[2]\cr}$\>{$\sqrt{n-1\over{2(n-2)^2}}$}\>{$~~~0$}\>{$~~~~0$}\>{~~~$\sqrt{n-3\over{2(n-2)}}$}\>{~~~~~~~~0}
\>{~~~~~~~~~~~0}\>{~~~~~~~-$\sqrt{(n-1)(n-3)\over{2(n-2)^2}}$}
\>{~~~~~~~~~~~~~~~~~0}\\
{}\\
$\matrix{
[21^2]~\cr
[21^{2}]^{*}\cr}$\>{~~~~~~$1$}\>{$~~~0$}\>{$~~~0$}\>{~~~~~~~~0}\>{~~~~~~~~~$0$}
\>{~~~~~~~~~~~~0}\>{~~~~~~~~~~~~~~~~0}\>{~~~~~~~~~~~~~~~~~0}\\
{}\\
$\matrix{
[21^2]\cr
[21]\cr}$\>{~~~~~~$0$}\>{$~\sqrt{3\over{8}}$}\>{~~~-${3\over{4}}$}\>{~~~~~~~~$0$}\>{~~~~~~~~-$1\over{4}$}
\>{~~~~~~~$~~~~0$}\>{~~~~~~~~~~~~~~0}\>{~~~~~~~~~~~~~~~~~0}\\
\>{}\\
$\matrix{
[211]~~\cr
[1^3]^{*2}\cr}$\>{~~~~~~$0$}\>{$~~~0$}\>{$~~~0$}\>{~~~~~~~~0}\>{~~~~~~~~~$1$}
\>{~~~~~~~~~~~~0}\>{~~~~~~~~~~~~~~~~0}\>{~~~~~~~~~~~~~~~~~0}\\
1111\=22222222222222\=33333333333\=44444444444\=5555555555\=666666666\=77777777\=88888888\=\kill\\
$\matrix{
[211]\cr
[1^2]\cr}$\>{$~~~~\sqrt{n+1\over{4(n-2)^2}}$}\>{$~~~0$}\>{$0$}\>{~~~0}\>{~~~0}\>{-$\sqrt{n-3\over{4(n-2)}}$}
\>{~~~$\sqrt{3(n-1)(n-3)\over{4(n-2)^2}}$}
\>{~~~~~~~~~~~~~0}\\
1111\=22222222222222\=3333333\=4444444444\=5555555555\=666666666\=77777777\=88888888\=\kill\\
$\matrix{
[2]\cr
[2]\cr}$\>{~~$\sqrt{(n-1)^2(n-3)\over{n(n-2)^2}}$}\>{$~~~0$}\>{~~~~~~~$0$}\>{~~~~~~-$\sqrt{n-1\over{n(n-2)}}$}
\>{~~~~~~~~~~~~0}\>{~~~~~~~~~~~~~~~0}\>{~~~~~~~~~~~~~$\sqrt{1\over{n(n-2)^2}}$}\>{~~~~~~~~~~~~~~~~~~~~~0}
\\
{}\\
$\matrix{
[2]\cr
[1]\cr}$\>{~~~~~~~~~~$0$}\>{$~~~0$}\>{$~~~\sqrt{n+1\over{4n}}$}\>{~~~~~~~~~$~~0$}\>{~~~~~~~~~$\sqrt{3(n-1)\over{4n}}$}
\>{~~~~~~~~~~~~~~~~0}\>{~~~~~~~~~~~~~~~~~~~~0}\>{~~~~~~~~~~~~~~~~-$\sqrt{1\over{2n}}$}\\
{}\\
$\matrix{
[2]\cr
[0]\cr}$\>{~~~~~~~~~~$0$}\>{$~~~0$}\>{$~~~~~~~0$}\>{~~~~~~~~~~~0}\>{~~~~~~~~~~~~$0$}
\>{~~~~~~~~~~~~~~~0}\>{~~~~~~~~~~~~~~~~~~~~1}\>{~~~~~~~~~~~~~~~~~~~~0}\\
{}\\
$\matrix{
[1^2]\cr
[1^2]\cr}$\>{~~$\sqrt{(n-1)^2(n-3)\over{(n+2)(n-2)^2}}$}\>{$~~~0$}\>{~~~~~~~$0$}\>{~~~~~~~~~~~~0}\>{~~~~~~~~~~~~~~0}
\>{~~~~~~~~~~~$\sqrt{n-1\over{n^2-4}}$}
\>{~~~~~~~~~~~~~-$\sqrt{3\over{(n+2)(n-2)^2}}$}\>{~~~~~~~~~~~~~~~~~~~~~0}
\\
{}\\
{$\matrix{
[1^2]\cr
[1]\cr}$}\>{~~~~~~~~~~0}\>{~~~~0}\>{~~~$\sqrt{3(n+1)\over{4(n+2)}}$}\>{~~~~~~~~~~~~0}\>{~~~~~~~~-$\sqrt{n-1\over{4(n-2)}}$}
\>{~~~~~~~~~~~~~~~~0}
\>{~~~~~~~~~~~~~~~~~~~~~0}
\>{~~~~~~~~~~~~~~~$\sqrt{3\over{2(n+2)}}$}\\
\=111111111111111111111111111111111111111111111111111111111111111111111111111111111111\=\kill\\
{--------------------------------------------------------------------------------------------------------------------------------------}\\
\end{tabbing}
\noindent $^{*}$ It can be taken as [211] or [21-1] when $n=7$.\\
\noindent $^{*2}$ It can be taken as [111] or [11-1] when $n=7$
\newpage
\hoffset=-3cm
\voffset=-1.cm

\noindent Table 9. ISFs $\left<
\matrix{
[\lambda_{1}] &[\lambda_{2}]\cr
[\nu_{1}] &[\nu_{2}]\cr}\right\vert\left.
\matrix{
[\lambda ]\cr
[\nu ]\cr}\right>$ of $SO(n)\supset SO(n-1)$ for $[20]\times [20]=[40]+[31]+$\\
\vskip .2cm
\noindent $[22]+[20]+[1^2]+[0]$ for $n\geq 5$.
\begin{tabbing}
\=111111111111111111111111111111111111111111111111111111111111111111111111111111111111\=\kill\\
{---------------------------------------------------------------------------------------------------------------------------------------------}\\
11111\=2222222222\=3333333333\=44444444444\=5555555555\=66666666666\=7777777777\=9999999999\=888888888\=\kill\\
{$\matrix{
[\lambda ]\cr
[\nu ]\cr}$ /}
\>{$\matrix{ [2] &[2]\cr
[2] &[2]\cr}$}\>{$\matrix{ [2] &[2]\cr
[2] &[1]\cr}$}\>{$\matrix{ [2] &[2]\cr
[2] &[0]\cr}$}\>{$\matrix{ [2] &[2]\cr
[1] &[2]\cr}$}
\>{$\matrix{ [2] &[2]\cr
[1] &[1]\cr}$}
\>{$\matrix{ [2] &[2]\cr
[1] &[0]\cr}$}\>{$\matrix{ [2] &[2]\cr
[0] &[2]\cr}$}\>{$\matrix{ [2] &[2]\cr
[0] &[1]\cr}$}\>{$\matrix{ [2] &[2]\cr
[0] &[0]\cr}$}\\
\=111111111111111111111111111111111111111111111111111111111111111111111111111111111111\=\kill\\
{---------------------------------------------------------------------------------------------------------------------------------------------}\\
11111\=2222222222\=3333333333\=44444444444\=5555555555\=66666666666\=7777777777\=9999999999\=888888888\=\kill\\
{$\matrix{
[4]\cr [4]\cr}$}\>{~~~~~~1}\>{~~~~~0}\>{~~~~~~0}\>{~~~~~0}\>{~~~~~0}\>{~~~~~0}\>{~~~~~0}\>{~~~~~0}\>{~~~~~0}\\
{}\\
{$\matrix{
[4]\cr
[3]\cr}$}\>{~~~~~~0}\>{~~~$\sqrt{1\over{2}}$}\>{~~~~~~0}\>{~~~~$\sqrt{1\over{2}}$}\>{~~~~~~0}\>{~~~~~~0}\>{~~~~~~0}\>{~~~~~~0}\>{~~~~~0}\\
{}\\
{$\matrix{
[4]\cr
[2]\cr}$}\>{$\sqrt{2(n-3)\over{3(n+4)(n-1)}}$}\>{~~~~~0}\>{-$\sqrt{n(n+3)\over{6(n+4)(n-1)}}$}
\>{~~~~~~~0}\>{~~$\sqrt{2(n+3)\over{3(n+4)}}$}\>{~~~~~~~0}\>{-$\sqrt{n(n+3)\over{6(n+4)(n-1)}}$}\>{~~~~~~~0}\>{~~~~~0}\\
{}\\
$\matrix{
[4]\cr
[1]\cr}$\>{~~~~~~0}\>{$\sqrt{n-2\over{(n+4)(n-1)}}$}\>{~~~~~~$0$}
\>{-$\sqrt{n-2\over{(n+4)(n-1)}}$}\>{~~~~~$0$}
\>{-$\sqrt{n(n+3)\over{2(n+1)(n+4)}}$}
\>{~~~~~~~~0}\>{-$\sqrt{n(n+1)\over{2(n-1)(n+4)}}$}\>{~~~~~0}\\
111\=222222222222\=333333333\=44444444\=5555555555\=666666666\=77777777\=888888888888\=99999999\=\kill\\
$\matrix{
[4]\cr
[0]\cr}$\>{$\sqrt{2(n-2)\over{(n+4)(n-1)(n+2)}}$}\>{$~~~~~~0$}\>{$~~~~~~~~0$}\>{~~~~~~~~~~~$~~0$}\>{~~~~~$\sqrt{4(n+1)\over{(n+2)(n+4)}}$}
\>{~~~~~~~~~~~~~~~~0}\>{~~~~~~~~~~~~~~~~~~~~0}\>{~~~~~~~~~~~~~~~0}\>{~~~~~~$\sqrt{n^2(n+1)\over{(n+4)(n+2)(n-1)}}$}\\
11111\=2222222222\=3333333333\=44444444444\=5555555555\=66666666666\=7777777777\=9999999999\=888888888\=\kill\\
{$\matrix{
[31]~\cr [31]^{*}\cr}$}\>{~~~~~~1}\>{~~~~~0}\>{~~~~~~0}\>{~~~~~0}\>{~~~~~0}\>{~~~~~0}\>{~~~~~0}\>{~~~~~0}\>{~~~~~0}\\
{}\\
{$\matrix{
[31]\cr
[3]\cr}$}\>{~~~~~~0}\>{~~~$\sqrt{1\over{2}}$}\>{~~~~~~0}\>{~~~-$\sqrt{1\over{2}}$}\>{~~~~~~0}\>{~~~~~~0}\>{~~~~~~0}\>{~~~~~~0}\>{~~~~~0}\\
{}\\
{$\matrix{
[31]\cr
[2]\cr}$}\>{~~~~~~0}\>{~~~~~~0}\>{~~~$\sqrt{1\over{2}}$}\>{~~~~~~0}\>{~~~~~~0}\>{~~~~~~0}\>{~~~~-$\sqrt{1\over{2}}$}\>{~~~~~~0}\>{~~~~~0}\\
{}\\
$\matrix{
[31]\cr
[1]\cr}$\>{~~~~~~0}\>{$\sqrt{n\over{(n+2)(n-1)}}$}\>{~~~~~~$0$}
\>{$\sqrt{n\over{(n+2)(n-1)}}$}\>{~~~~~$0$}
\>{-$\sqrt{(n-2)(n+1)\over{2(n-1)(n+2)}}$}
\>{~~~~~~~~0}\>{$\sqrt{(n-2)(n+1)\over{2(n-1)(n+2)}}$}\>{~~~~~0}\\
{}\\
$\matrix{
[31]~~\cr
[21]^{*2}\cr}$\>{~~~~~~0}\>{~~~$\sqrt{1\over{2}}$}\>{~~~~~~0}\>{~~~-$\sqrt{1\over{2}}$}\>{~~~~~~0}\>{~~~~~~0}\>{~~~~~~~~0}\>{~~~~~~0}\>{~~~~~0}\\
\=111111111111111111111111111111111111111111111111111111111111111111111111111111111111\=\kill\\
{---------------------------------------------------------------------------------------------------------------------------------------------}\\
\end{tabbing}
\vskip -1.cm
\noindent $^{*}$ It can be taken as [31] or [3-1] when $n=5$.\\
\noindent $^{*2}$ It can be taken as [21] or [2-1] when $n=5$
\newpage

\hoffset=-3cm
\noindent Table 9. (Continued)
\begin{tabbing}
\=111111111111111111111111111111111111111111111111111111111111111111111111111111111111\=\kill\\
{---------------------------------------------------------------------------------------------------------------------------------------------}\\
11111\=2222222222\=3333333333\=44444444444\=5555555555\=66666666666\=7777777777\=9999999999\=888888888\=\kill\\
{$\matrix{
[\lambda ]\cr
[\nu ]\cr}$ /}
\>{$\matrix{ [2] &[2]\cr
[2] &[2]\cr}$}\>{$\matrix{ [2] &[2]\cr
[2] &[1]\cr}$}\>{$\matrix{ [2] &[2]\cr
[2] &[0]\cr}$}\>{$\matrix{ [2] &[2]\cr
[1] &[2]\cr}$}
\>{$\matrix{ [2] &[2]\cr
[1] &[1]\cr}$}
\>{$\matrix{ [2] &[2]\cr
[1] &[0]\cr}$}\>{$\matrix{ [2] &[2]\cr
[0] &[2]\cr}$}\>{$\matrix{ [2] &[2]\cr
[0] &[1]\cr}$}\>{$\matrix{ [2] &[2]\cr
[0] &[0]\cr}$}\\
\=111111111111111111111111111111111111111111111111111111111111111111111111111111111111\=\kill\\
{---------------------------------------------------------------------------------------------------------------------------------------------}\\
11111\=2222222222\=3333333333\=44444444444\=5555555555\=66666666666\=7777777777\=9999999999\=888888888\=\kill\\
{$\matrix{
[31]~~\cr
[1^2]^{*3}\cr}$}\>{$~\sqrt{1\over{n+2}}$}\>{~~~~~0}\>{~~~~~~0}\>{~~~~~0}\>{~~$\sqrt{n+1\over{n+2}}$}\>{~~~~~0}\>{~~~~~0}\>{~~~~~0}\>{~~~~~0}\\
{}\\ {$\matrix{
[22]~~\cr
[22]^{*4}\cr}$}\>{~~~~1}\>{~~~$~~~0$}\>{~~~~~~0}\>{~~~~~~0}\>{~~~~~~0}\>{~~~~~~0}\>{~~~~~~0}\>{~~~~~~0}\>{~~~~~0}\\
{}\\
{$\matrix{
[22]~~\cr
[21]^{*2}\cr}$}\>{~~~~~0}\>{~~~$\sqrt{1\over{2}}$}\>{~~~~~~0}\>{~~~$\sqrt{1\over{2}}$}\>{~~~~~~0}\>{~~~~~~0}\>{~~~~~~0}\>{~~~~~~0}\>{~~~~~0}\\
{}\\
{$\matrix{
[22]\cr
[2]\cr}$}\>{$\sqrt{n+3\over{3(n-2)(n-1)}}$}\>{~~~~~0}\>{$\sqrt{n(n-3)\over{3(n-2)(n-1)}}$}
\>{~~~~~~~0}\>{~~$\sqrt{n-3\over{3(n-2)}}$}\>{~~~~~~~0}\>{$\sqrt{n(n-3)\over{3(n-2)(n-1)}}$}\>{~~~~~~~0}\>{~~~~~0}\\
111\=222222222222\=33333333\=4444444444444\=55555555\=6666666666666\=77777777\=999999999999\=888888888\=\kill\\
{$\matrix{
[2]\cr
[2]\cr}$}\>{$\sqrt{n(n^2-9)\over{(n+4)(n-2)(n-1)}}$}\>{~~~~~0}\>{-$\sqrt{4\over{(n+4)(n-2)(n-1)}}$}
\>{~~~~~~~0}\>{~-$\sqrt{n\over{(n+4)(n-2)}}$}\>{~~~~~~~0}\>{-$\sqrt{4\over{(n+4)(n-2)(n-1)}}$}\>{~~~~~~~0}\>{~~~~~0}\\
11111\=222222222\=33333333333\=44444444444\=5555555555\=66666666666\=7777777777\=9999999999\=888888888\=\kill\\
$\matrix{
[2]\cr
[1]\cr}$\>{~~~~~~0}\>{$\sqrt{n(n+1)\over{2(n+4)(n-1)}}$}\>{~~~~~~~$0$}\>{-$\sqrt{n(n+1)\over{2(n+4)(n-1)}}$}\>{~~~~~$0$}
\>{$\sqrt{n-2\over{(n-1)(n+4)}}$}
\>{~~~~~~~~0}\>{$\sqrt{n-2\over{(n-1)(n+4)}}$}\>{~~~~~0}\\
111\=222222222222\=333333333\=44444444\=5555555555\=666666666\=77777777\=888888888888\=99999999\=\kill\\
$\matrix{
[2]\cr
[0]\cr}$\>{$~\sqrt{2(n+1)\over{(n+4)(n-1)}}$}\>{$~~~~~~0$}\>{$~~~~~~~~0$}\>{~~~~~~~~~~~$~~0$}\>{~~~~~~~~$\sqrt{n-2\over{n+4}}$}
\>{~~~~~~~~~~~~~~~~0}\>{~~~~~~~~~~~~~~~~~~~~0}\>{~~~~~~~~~~~~~~~0}\>{~~~~~~~-$\sqrt{4(n-2)\over{(n+4)(n-1)}}$}\\
11111\=2222222222\=3333333333\=44444444444\=5555555555\=66666666666\=7777777777\=9999999999\=888888888\=\kill\\
{$\matrix{
[1^2]~~\cr
[1^2]^{*3}\cr}$}\>{~~$\sqrt{n+1\over{n+2}}$}\>{~~~~~0}\>{~~~~~~0}\>{~~~~~0}\>{~-$\sqrt{1\over{n+2}}$}\>{~~~~~0}\>{~~~~~0}\>{~~~~~0}\>{~~~~~0}\\
{}\\ 
$\matrix{
[1^2]\cr
[1]\cr}$\>{~~~~~~0}\>{$\sqrt{(n+1)(n-2)\over{2(n+2)(n-1)}}$}\>{~~~~~~$0$}\>{$\sqrt{(n+1)(n-2)\over{2(n+2)(n-1)}}$}\>{~~~~~$0$}
\>{$\sqrt{n\over{(n-1)(n+2)}}$}
\>{~~~~~~~~0}\>{-$\sqrt{n\over{(n-1)(n+2)}}$}\>{~~~~~0}\\
111\=222222222222\=333333333\=44444444\=5555555555\=666666666\=77777777\=888888888888\=99999999\=\kill\\
$\matrix{
[0]\cr
[0]\cr}$\>{~$\sqrt{(n+1)(n-2)\over{(n-1)(n+2)}}$}\>{$~~~~~~0$}\>{$~~~~~~~~0$}\>{~~~~~~~~~~~$~~0$}\>{~~~~~~~-$\sqrt{n\over{n+2}}$}
\>{~~~~~~~~~~~~~~~~0}\>{~~~~~~~~~~~~~~~~~~~~0}\>{~~~~~~~~~~~~~~~0}\>{~~~~~~~~$\sqrt{2\over{(n+2)(n-1)}}$}\\
\=111111111111111111111111111111111111111111111111111111111111111111111111111111111111\=\kill\\
{----------------------------------------------------------------------------------------------------------------------------------------------}\\
\end{tabbing}
\noindent $^{*3}$ It can be taken as [11] or [1-1] when $n=5$.\\
\noindent $^{*4}$ It can be taken as [22] or [2-2] when $n=5$

\end{document}